\documentclass[twocolumn]{aastex631}
\received{13-Dec-2021}
\revised{20-May-2022}
\accepted{25-May-2022}
\submitjournal{ApJS}

\shorttitle{B/T Ratio and sSFR Enhancement of Paired Galaxies}
\shortauthors{He et al.}

\begin{document}

\title{Close Major Merger Pairs at $z=0$:\\
Bulge-to-Total Ratio and Star Formation Enhancement}

\correspondingauthor{Cong Kevin Xu}
\email{congxu@nao.cas.cn}

\author[0000-0003-1761-5442]{Chuan He}
\affiliation{Chinese Academy of Sciences South America Center for Astronomy, National Astronomical Observatories of China, Chinese Academy of Sciences, Beijing 100101, PR China}
\affiliation{School of Astronomy and Space Sciences, University of Chinese Academy of Sciences, Beijing 100049, PR China}

\author[0000-0002-1588-6700]{Cong Kevin Xu}
\affiliation{Chinese Academy of Sciences South America Center for Astronomy, National Astronomical Observatories of China, Chinese Academy of Sciences, Beijing 100101, PR China}

\author[0000-0001-5662-7169]{Donovan Domingue}
\affiliation{Georgia College \& State University, CBX 82, Milledgeville, GA 31061, USA}

\author[0000-0002-6166-3647]{Chen Cao}
\affiliation{School of Space Science and Physics, Shandong University, Weihai, Shandong 264209, PR China}
\affiliation{Shandong Key Laboratory of Optical Astronomy \& Solar-Terrestrial Environment, Institute of Space Sciences, Shandong University, Weihai, Shandong 264209, PR China}

\author[0000-0001-6511-8745]{Jia-sheng Huang}
\affiliation{Chinese Academy of Sciences South America Center for Astronomy, National Astronomical Observatories of China, Chinese Academy of Sciences, Beijing 100101, PR China}

\begin{abstract}
We present a study of the bulge-to-total ratio (B/T) of a Ks-band-selected sample of 88 close major-merger pairs of galaxies (H-KPAIR) based on 2-D decomposition of SDSS r-band images with \textsc{galfit}. We investigate the dependence of the interaction-induced specific star formation rate enhancement ($\rm sSFR_{enh}$) on the B/T ratio, and the effects of this dependence on the differences between star-forming galaxies (SFGs) in spiral+spiral (S+S) and spiral+elliptical (S+E) pairs. Of all 132 spiral galaxies in H-KPAIR, the 44 in S+E pairs show higher B/T than those in the 44 S+S pairs, with means of $\rm B/T = 0.35 \pm 0.05$ and $\rm B/T = 0.26 \pm 0.03$, respectively. There is a strong negative dependence of $\rm sSFR_{enh}$ on the B/T ratio and only paired SFGs with $\rm B/T<0.3$ show significant ($>5\sigma$) enhancement. Paired SFGs in S+S pairs show a similar trend, and many disky SFGs ($\rm B/T<0.1$) in S+S have strong sSFR enhancements ($\rm sSFR_{enh} > 0.7$~dex). For SFGs in S+E, the sSFR has no clear B/T dependence, nor any significant enhancement in any B/T bin. Disky SFGs in S+S show significant ($>4\sigma$) enhancement in the molecular gas content ($\rm M_{H_2}/M_{star}$), while SFGs in S+E have no such enhancement in any B/T bin. No significant enhancement on total gas content ($\rm M_{gas}/M_{star}$) is found in any B/T bin for paired galaxies. The star formation efficiency of either the total gas ($\rm SFE_{gas} = SFR/M_{gas}$) or the molecular gas ($\rm SFE_{H_2} = SFR/M_{H_2}$) does not depend on the B/T ratio. The only significant ($>4\sigma$) SFE enhancement found for paired SFGs is the $\rm SFE_{gas}$ for disky SFGs in S+S pairs.
\end{abstract}

\keywords{Galaxy evolution (594), Galaxy interactions (600), Galaxy structure (622), Star formation (1569)}

\section{Introduction} \label{sec:intro}

It has been well established that galaxy interactions and mergers can induce star formation enhancement \citep{*1972ApJ...178..623T,*1978ApJ...219...46L,1987AJ.....93.1011K,*1996ARA&A..34..749S}. In the local universe, the most extreme starbursts such as the ultraluminous infrared galaxies ($\rm ULIRGs: L_{IR} \geq 10^{12} L_\sun$) are exclusively found in the final stage of mergers \citep*{1996ARA&A..34..749S}. Significant star formation enhancements are also detected in interacting galaxies in earlier merger stages such as those in optically selected pairs \citep{1987AJ.....93.1011K,2000ApJ...530..660B,*1991ApJ...374..407X,2004MNRAS.355..874N,2010MNRAS.407.1514E,2012MNRAS.426..549S}. Statistical studies based on large surveys found that, among early stage mergers, star-forming galaxies (SFGs) in close major-merger pairs (separation $\rm \lesssim 30\;kpc$ and mass-ratio $\lesssim 3$) have the highest star formation rate (SFR) enhancement \citep{*1991ApJ...374..407X,2012MNRAS.426..549S,2013MNRAS.433L..59P}. However, Spitzer observations of a sample of Ks-band-selected close major-merger pairs \citep{2010ApJ...713..330X} found that only $\sim 25\%$ of SFGs in the sample show strong enhancement in specific star formation rate ($\rm sSFR=SFR/M_{star}$). Furthermore, the far-infrared (FIR) observations by Spitzer and Herschel show that only SFGs in spiral-spiral (here after S+S) pairs have significantly enhanced sSFR, but not those in spiral-elliptical (here after S+E) pairs \citep{2010ApJ...713..330X,2016ApJS..222...16C,2016ApJ...829...78D}. The GBT HI observations of \citet{2018ApJS..237....2Z} and IRAM CO observations of \citet{2019A&A...627A.107L} for paired galaxies, selected from H-KPAIR sample of 88 close major-merger pairs that have Herschel FIR observations \citep{2016ApJS..222...16C}, found no significant difference between the total gas abundances of SFGs in S+E and in S+S pairs. These results reject the hypothesis that the lack of star formation enhancement in S+E pairs is due to stripping of cold gas of the spiral component by ram-pressure of the hot-gas halo surrounding the elliptical component \citep{2009ApJ...691.1828P,2011A&A...535A..60H}.

Simulations of interacting galaxies have shown that a massive bulge can stabilize the disk and suppress the SFR during and after close encounters \citep{*1996ApJ...464..641M,2008A&A...492...31D,2008MNRAS.384..386C}. This mechanism may play an important role in the low frequency of star formation enhancement in paired galaxies, in particular those in S+E pairs. However, there has been no observational test of the theoretical prediction in the literature. Indeed, such a test requires accurate estimates of the bulge-to-total (B/T) ratios of paired galaxies. Many works have been carried out on the task of decompositing large samples of galaxies with image fits \citep[such as][]{2011ApJS..196...11S,*2012MNRAS.421.2277L,2015MNRAS.446.3943M,2016MNRAS.455.2440M,2016ApJS..225....6K}. \citet{2011ApJS..196...11S} presented a two-dimensional, point-spread-function-convolved, bulge-disk decompositions in the g and r bands on a sample of 1.1 million SDSS galaxies with \textsc{gim2d} \citep{2002ApJS..142....1S}. \citet*{2012MNRAS.421.2277L} made efforts on low redshift ($\rm z<0.05$) galaxies with their own pipeline. \citet{2015MNRAS.446.3943M} carried out decompositions in the r-band on 670,722 SDSS spectroscopic galaxies, a subsample of \citet{2011ApJS..196...11S}, with \textsc{galfit} \citep{2002AJ....124..266P}, and extended it to the g and i bands in \citet{2016MNRAS.455.2440M}. \citet{2016ApJS..225....6K} provided a recipe for choosing the initial guess values of the input parameters based on galaxy color. However, these works are automated on bulks of sample without individual inspection, therefore, their results on interacting and distorted galaxies may not be reliable.

In this paper, we present our own bulge-disk decomposition for galaxies in the H-KPAIR sample \citep{2016ApJS..222...16C}, based on \textsc{galfit} and manual intervention deblending photometry. We study the effects of the central bulge on the star formation enhancement in paired galaxies. Using the H-KPAIR sample and a well matched control sample, we quantify these effects and examine how much the sSFR difference between S+E and S+S pairs is caused by them. The H-KPAIR sample and control sample are introduced in Section \ref{sec:sample} and \ref{sec:ctrl}, respectively. In Section \ref{sec:bt} we describe how we measure the B/T of H-KPAIR galaxies and compare our results with those in the literature. The science analyses are presented in Section \ref{sec:result}, followed by a discussion in Section \ref{sec:discussion} and conclusions in Section \ref{sec:conclusion}. Throughout this paper, we adopt the $\Lambda$-cosmology with $\rm \Omega_m=0.3$ and $\Omega_\Lambda=0.7$, and $\rm H_0=70\;km\;s^{-1}\;Mpc^{-1}$.

\section{The Local Close Major-Merger Sample (H-KPAIR)} \label{sec:sample}

Our pair sample is identical with the local close major-merger sample H-KPAIR \citep{2016ApJS..222...16C}, which is a subsample of a complete and unbiased Ks-band \citep[Two Micron All Sky Survey, 2MASS,][]{2006AJ....131.1163S,2000AJ....119.2498J,2MASS_XSC_data} selected sample KPAIR \citep{2009ApJ...695.1559D}. All H-KPAIR galaxies have spectroscopic redshifts in the range of $0.0067<z<0.1$. The pair sample requires the projected separations range of $\rm 5\;h^{-1}\;kpc \leq s(p) \leq 20\;h^{-1}\;kpc$, the radial relative velocity $\rm \delta(Vz) < 500\;km\;s^{-1}$ and the Ks-band magnitude differences within 1 mag (corresponding to a mass ratio no greater than 2.5). H-KPAIR sample contains 44 S+S pairs and 44 S+E pairs, all have Herschel imaging observations in the 6 bands at 70, 110, 160, 250, 350, and $500\;\micron$ \citep{2016ApJS..222...16C}. Furthermore, in H-KPAIR, 70 pairs have single dish 21cm HI observations \citep{2018ApJS..237....2Z} and 78 S galaxies are observed by the IRAM 30m telescope for CO emissions \citep{2019A&A...627A.107L}.

We adopt SFR of H-KPAIR galaxies in \citet{2016ApJS..222...16C}, which are derived from $\rm L_{IR}$ \citep*[8–1000 \micron;][]{1996ARA&A..34..749S} using the formula of \citet{1998ApJ...498..541K}, with an additional correction factor of $10^{-0.20}$ for the conversion from the Salpeter IMF to the Kroupa IMF \citep{2013seg..book..419C}, where the $\rm L_{IR}$ is generated from SED fits with the dust emission model of \citet*{2007ApJ...657..810D}

\begin{figure}[htb!]
\includegraphics[width=0.47\textwidth]{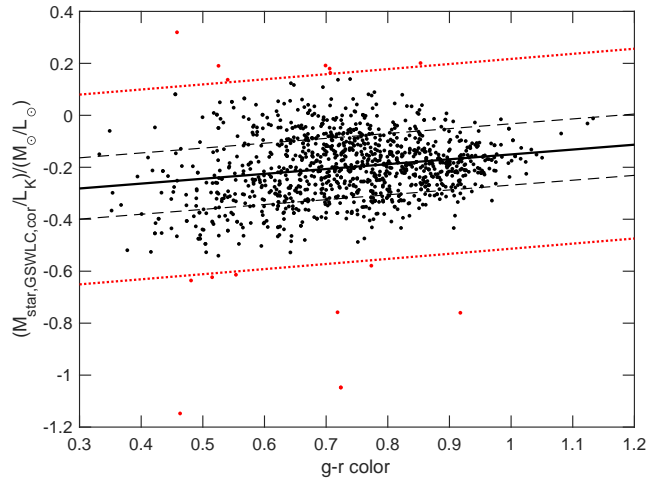}
\caption{Correlation between mass-to-light ratio $\rm M_{star}/L_K$ and g-r color. The black solid line represents the linear regression and the black dashed lines the 1$\sigma$ deviation. The red dotted lines show the 3$\sigma$ boundaries and the red dots the outliers excluded from the fits.
\label{fig:g-r}}
\end{figure}

In \citet{2016ApJS..222...16C} the stellar mass $\rm M_{star}$ of H-KPAIR galaxies was estimated from the Ks-band luminosity using a constant mass-to-light ratio $\rm M_{star}/L_K$ \citep{2004ApJ...603L..73X,2009ApJ...695.1559D,2012ApJ...747...85X}. This is because for normal galaxies both the near infrared (NIR) emission and the stellar mass are dominated by the old stellar populations and therefore the  $\rm M_{star}/L_K$ ratio is nearly independent of the galaxy types \citep{1996A&A...312..397G}. On the other hand \citet{2001ApJ...550..212B} showed a color dependence of the $\rm M_{star}/L_K$ ratio caused by variations of the star formation history, although it is significantly weaker than that for the  mass-to-light ratio in optical bands.  In this paper, we improve the stellar mass estimate by including a g-r color dependence in the $\rm M_{star}/L_{K}$ ratio. To derive the g-r color for a galaxy, we make $\rm 25\;mag\;arcsec^{-2}$ isophotal photometry using \textsc{se}xtractor \citep{1996A&AS..117..393B} on the SDSS r-band image and apply the same isophotal aperture to the g-band image. For H-KPAIR sample, we visually check and optimize the segmentations for all the pairs. When \textsc{se}xtractor fails to make good deblending for a merger, we manually draw polygon apertures for them and use the $\rm 25\;mag\;arcsec^{-2}$ isophot of the whole merger to determine the outer boundaries of the apertures for both galaxies. We exploit the stellar mass derived in GALEX–SDSS–WISE LEGACY CATALOG \citep[GSWLC;][]{2016ApJS..227....2S,2018ApJ...859...11S} for the calibration of the relation between $\rm M_{star}/L_K$ and g-r. Among 1320 of our control galaxies, 1180 have $\rm M_{star}$ measurements in the GSWLC-2 \citep{2016ApJS..227....2S,2018ApJ...859...11S}. For them we carry out a linear regression between the $\rm M_{star,GSWLC,cor}/L_K$ and the g-r color, where the $\rm M_{star,GSWLC,cor}$ is the stellar mass in GSWLC-2 after correcting the difference between the cosmology parameters used in GSWLC-2 and in this paper. After excluding the 3$\sigma$ outliers iteratively and assuming $\rm M_{star,GSWLC,cor}$ is a good $\rm M_{star}$ estimator, we obtain the following relation: 
\begin{equation}
\rm M_{star}/L_K = 0.19 \times (g-r) - 0.34 . 
\label{eq:m/l}
\end{equation}
The result is shown in Figure \ref{fig:g-r}. The $\rm M_{star}$ of galaxies in our samples are calculated using the formula in Equation \ref{eq:m/l}.

\section{Two-Component Model fits (\textsc{galfit}) of H-KPAIR} \label{sec:bt}
\setcounter{footnote}{5}
We carry out two-component two-dimensional fits on SDSS r-band images with \textsc{galfit}\citep{2002AJ....124..266P}. The S\'ersic \citeyearpar{1968adga.book.....S} and Exponential \citep{1970ApJ...160..811F} light profiles are adopted for the bulge and the disk, respectively. This method has been well tested \citep[e.g.,][]{2015MNRAS.446.3943M,2016ApJS..225....6K} on the task of decomposing the bulge and the disk of SDSS local galaxies. We constrain the S\'ersic index of the ``bulge" component in the range of $\rm 1 \leq n \leq 8$.

Because some of our galaxies are on the edge of the field, or even be cut into multiple parts found in different fields, we use the SDSS SAS mosaic tool\footnote{\url{dr12.sdss.org/mosaics/}}, which can stitch together several sky-subtracted, calibrated frames\footnote{\url{data.sdss.org/datamodel/files/BOSS_PHOTOOBJ/frames/RERUN/RUN/CAMCOL/frame.html}} to form a coherent image over a specify patch of sky using the \textsc{sw}arp \citep{2002ASPC..281..228B}. According to \citet{2015MNRAS.446.3943M}, the image should have at least 20 half-light radii to provide enough pixels for background. We use images of $909\times909$ pixels uniformly for the fits of nearly all of our sample galaxies, corresponding to $0\fdg1$ square sky, since the half-light radii of them are no more than 45 pixels derived from \textsc{se}xtractor. The only exception is galaxy J20471908+0019150 (of pair J2047+0018), the largest galaxy in our sample (size $\sim 3\arcmin$), for which an image of $1818\times1818$ pixels is used.

The image mask is generated based on the \textsc{se}xtractor segmentation image. We first run \textsc{se}xtractor with the detect threshold set at $\rm 25\;mag\;arcsec^{-2}$ to detect all the ``source" out of sky background. Then we identify our target galaxies visually and set segmentation area of other sources for mask. We also edit mask areas manually when \textsc{se}xtractor fails to deblend the sources.

The point-spread-function (psf) is generated from the PsField files\footnote{\url{data.sdss.org/datamodel/files/PHOTO\_REDUX/RERUN/RUN/objcs/CAMCOL/psField.html}} using the code read\textsc{a}tlas\textsc{i}mages-v5\_4\_11 provided by SDSS\footnote{\url{classic.sdss.org/dr7/products/images/read\_psf.html}}.

We calculate an equivalent gain $\rm GAIN_{eq}$ and let \textsc{galfit} generate the Poisson-noise sigma image itself. $\rm GAIN_{eq}$ is defined as $\rm GAIN_{eq}=GAIN/cimg$, where GAIN is the original CCD gain and cimg the calibration factor from DN into nanomaggie, both are listed in the SDSS website\footnote{\url{data.sdss.org/datamodel/files/BOSS_PHOTOOBJ/frames/RERUN/RUN/CAMCOL/frame.html}}.

\begin{figure*}[htb!]
\includegraphics[width=1\textwidth]{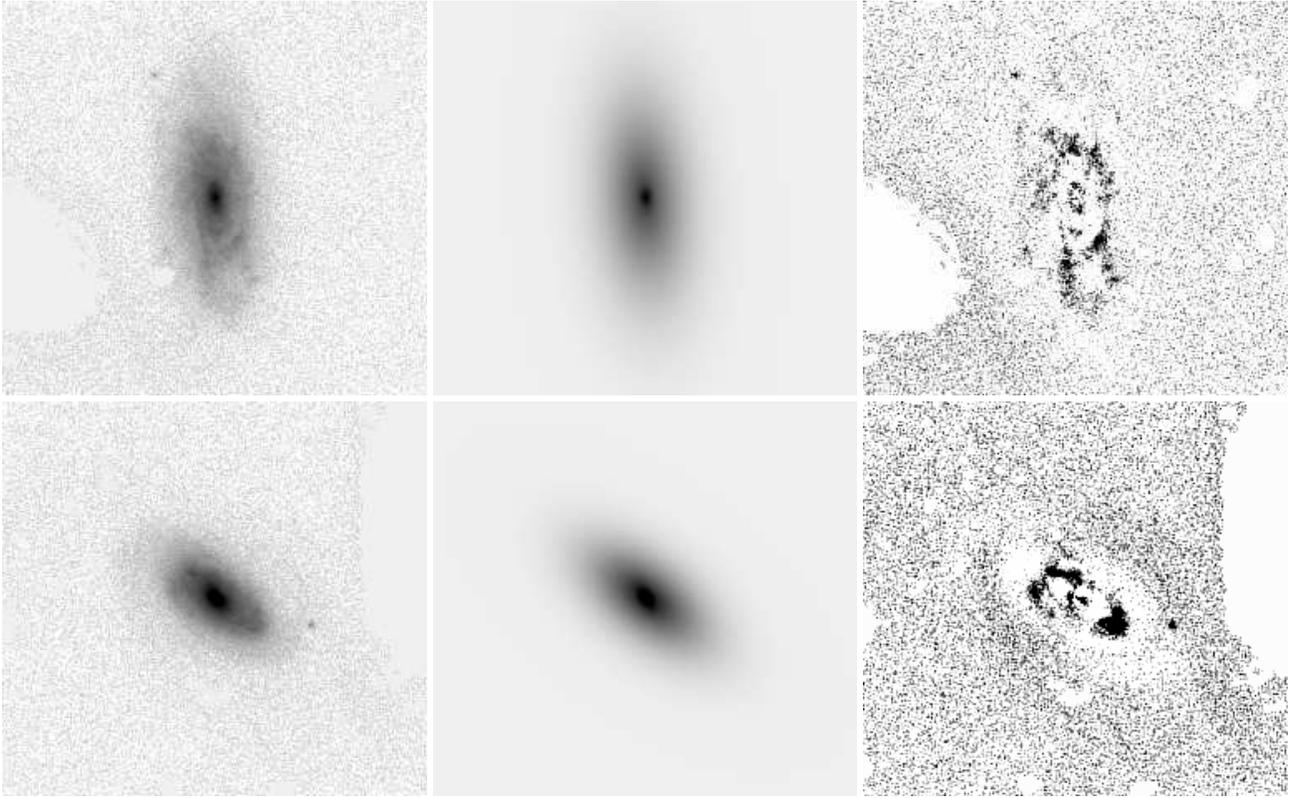}
\caption{\textsc{galfit} results of the pair J0829+5531 (upper panels: galaxy J08291491+5531227; lower panels: galaxy J08292083+5531081; left: SDSS r-band image; middle: model image; right: residual). This is an example of satisfactory fits using S\'ersic-bulge and disk models, although some substructures such as spirals and bars are left in the residual because of their neglect in the models.
\label{fig:show_2comp}}
\end{figure*}

\begin{figure*}[htb!]
\includegraphics[width=1\textwidth]{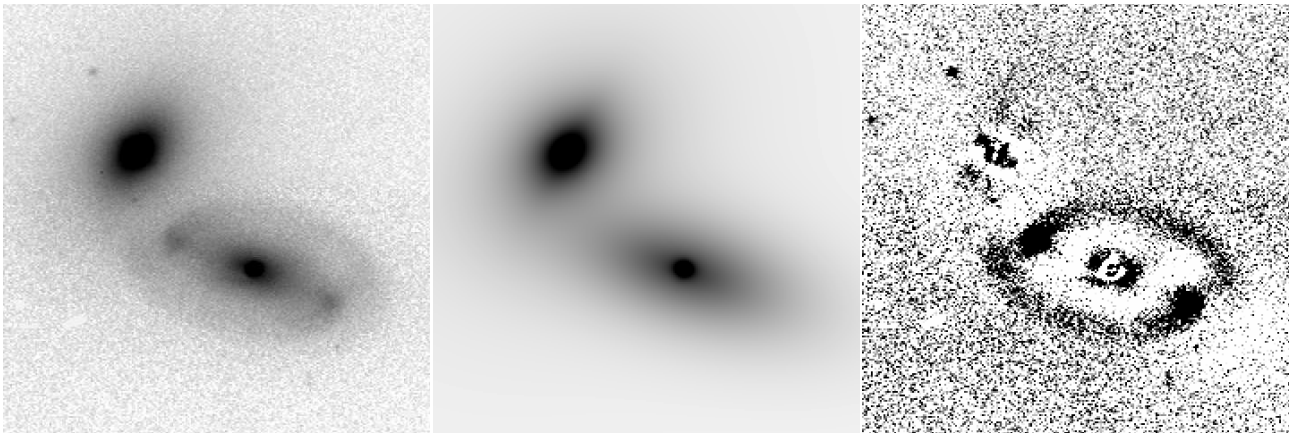}
\caption{\textsc{galfit} results of the pair J0020+0049 (left: SDSS r-band image; middle: model image; right: residual). This is an example of simultaneous fits of both galaxies in a pair using two S\'ersic-bulge + disk models.
\label{fig:show_4comp}}
\end{figure*}

\begin{figure*}[htb!]
{
\fig{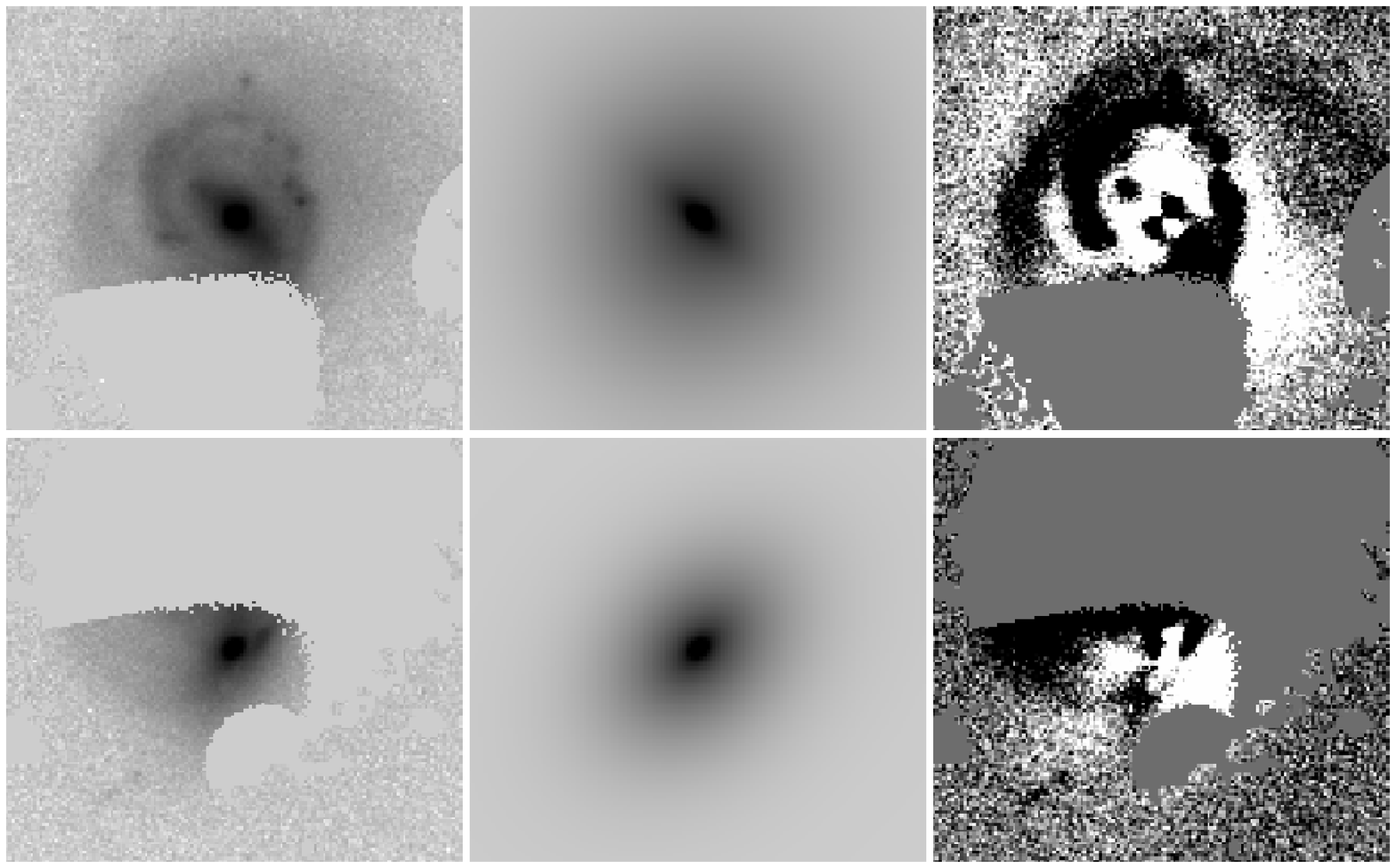}{0.6\textwidth}{(a)}
\fig{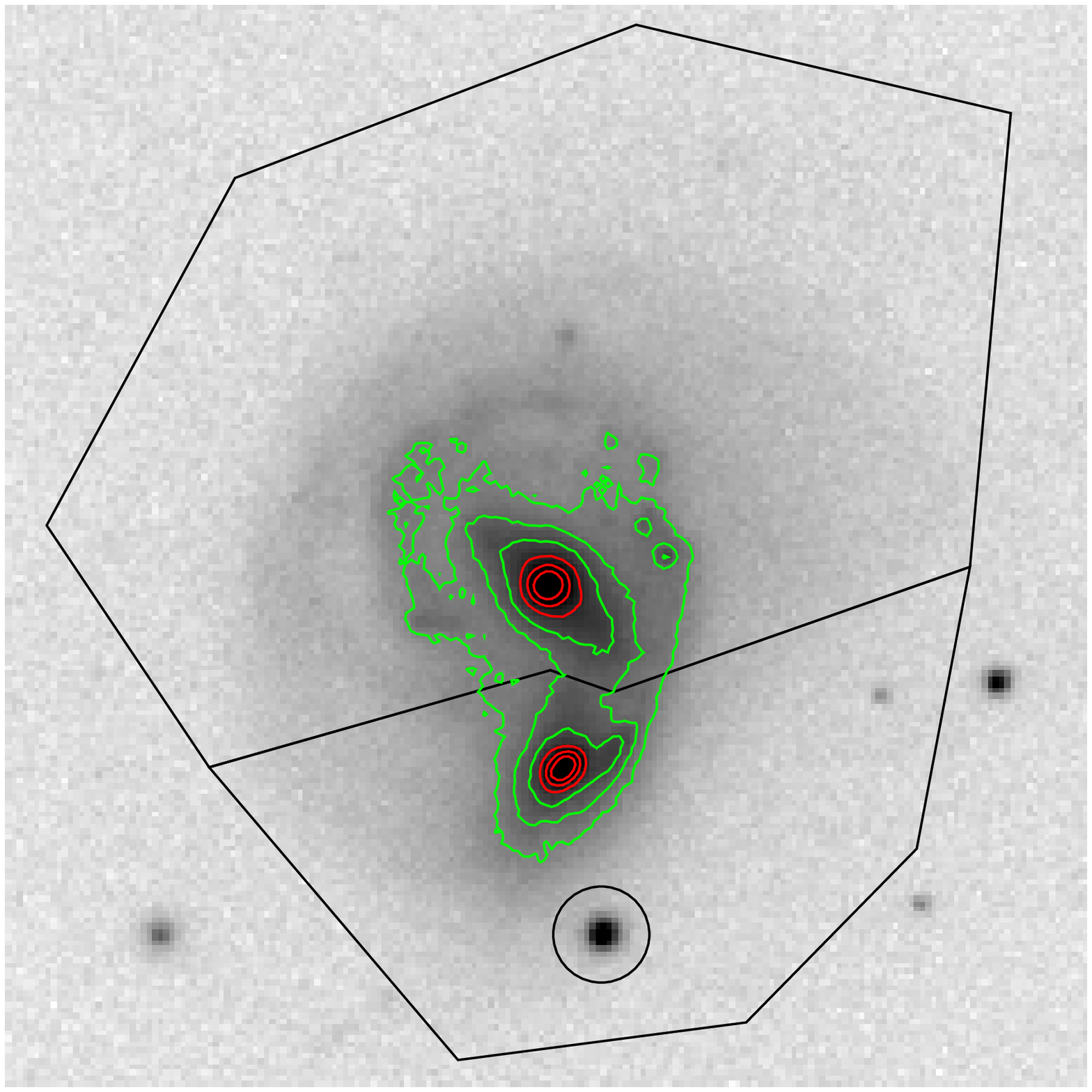}{0.37\textwidth}{(b)}
}
% \gridline{
% }
\caption{(a) \textsc{galfit} results of the merging pair J1444+1207 (top panels: J14442079+1207552; bottom panels: J14442055+1207429; left: SDSS r-band image; middle: model image; right: residual). (b) The polygon apertures (black) for the photometry on total fluxes of the two galaxies, and isophotal contours of the pair. The red contour levels are at 18.5, 19, 19.5, and the green contour levels are at 20, 20.5, 21 $\rm mag\;arcsec^{-2}$ (no smoothing). The most inner isophot is still larger than the FWHM of the psf. This is an example of bulge fits and deblending photometry on distorted galaxies. The compact, undistorted, high S\'ersic index component—bulge—is constrained well by the brightest pixels on the galaxy center, and is fitted well, although an exponential ellipse cannot represent the distorted disk reasonably. 
\label{fig:show_poly}}
\end{figure*}

The standard \textsc{galfit} two-component fits is carried out for most H-KPAIR galaxies. Simultaneous fits of both galaxies in a pair are carried out for 61 close pairs, and galaxies in the other 16 well separate pairs are fitted individually. These fits yield reasonably good results, as illustrated by the examples shown in Figure~\ref{fig:show_2comp} and Figure~\ref{fig:show_4comp}. However, the standard \textsc{galfit} process fails to work for several extremely distorted and merging pairs, either producing large chi-squares or not converging at all. For these cases (11 merging pairs), the following special procedure is performed. Firstly, we assume that bulges in these systems can still be well fitted by 2-D models. This is because, compared to the disks, bulges which are dynamically ``hot" respond to tidal interactions more subtly. The triggered morphological distortions such as surface-brightness excess in the outer regions and slightly off-enteric inner isophotes \citep{1977ApJ...218..333K,1986ApJ...307...97A,1988A&A...201L..30D,2019MNRAS.488..830M} shall have minimal effect on the 2-D model-fits for the estimate of the bulge luminosity. Accordingly, for each system, \textsc{galfit} is carried out in order to obtain the model fluxes of bulges while neglecting the goodness of the fits of the disks. And then, secondly, the total flux of each galaxy in a pair is measured by deblending photometry with \textsc{se}xtractor $\rm 25\;mag\;arcsec^{-2}$ or hand-drawn polygon apertures (see Section \ref{sec:sample}). An example is shown in Figure~\ref{fig:show_poly}). The central isophotes (the red contours) of galaxies in this example show that the bulges are not strongly distorted. Nevertheless, the B/T ratios so obtained may have larger uncertainties than those from ordinary \textsc{galfit}. Indeed, among the 11 merging pairs, three galaxies have the model bulge flux larger than the total flux from polygon aperture photometry. Since these are likely bulge-dominated galaxies, they are regarded as having $\rm B/T=1$.

The \textsc{galfit} results are listed in Table \ref{tab:catalog}. Note that for pairs with the two galaxies fitted simultaneously, the reduced chi-square $\chi^2/\nu$ is listed in the row of the first galaxy of each pair. For those only the bulge is fitted and the total flux obtained from $\rm 25\;mag\;arcsec^{-2}$ isophot or polygon aperture (with flag bit 2), the $\rm m_{GAL}$ is taken from the photometry, and the disk magnitude $\rm m_D$ is calculated by subtracting the bulge flux from the total flux. Other disk parameters of these galaxies are not available, and their $\chi^2/\nu$ are not listed. We note that, even though \textsc{galfit} exploits psf deconvolved images, any model component with radius significantly below the seeing (with a median FWHM of $1\farcs32$) shall be taken with caution.

\clearpage
\startlongtable
\begin{deluxetable*}{lrcrrrrrrrrrrrrrc}
\tablecaption{\textsc{galfit} Results of H$-$KPAIR\label{tab:catalog}}
\tablewidth{0pt}
\tabletypesize{\scriptsize}
%\footnotesize for 10 pt, \small for 11 pt, \scriptsize for 12 pt (default)
\tablehead{\colhead{ID} & \colhead{z} & \colhead{Type} & \colhead{$\rm m_{iso}$} & \colhead{$\rm m_{GAL}$} & \colhead{B/T} & \colhead{${\rm m}_B$} & \colhead{${\rm n}_B$} & \colhead{re} & \colhead{${\rm (b/a)}_B$} & \colhead{${\rm PA}_B$} & \colhead{${\rm m}_D$} & \colhead{rs} & \colhead{${\rm (b/a)}_D$} & \colhead{${\rm PA}_D$} & \colhead{$\rm \chi^2/\nu$} & \colhead{flag}\\
\colhead{} & \colhead{} & \colhead{} & \colhead{(mag)} & \colhead{(mag)} & \colhead{} & \colhead{(mag)} & \colhead{} & \colhead{(arcs)} & \colhead{} & \colhead{(deg)} & \colhead{(mag)} & \colhead{(arcs)} & \colhead{} & \colhead{(deg)} & \colhead{} & \colhead{}}
\decimalcolnumbers
\startdata
J00202580$+$0049350 & 0.0149 & S,J & 13.66 & 13.72 & 0.11 & 16.12 & 1.58 & 0.9 & 0.71 & 76.6 & 13.85 & 11.5 & 0.45 & 72.7 & 1.12 & 1 \\
J00202748$+$0050009 & 0.0162 & E,J & 13.39 & 13.01 & 0.97 & 13.05 & 6.78 & 16.3 & 0.77 & $-$49.7 & 16.74 & 4.6 & 0.34 & $-$30.7 & \nodata & 1 \\
J01183417$-$0013416 & 0.0453 & S,J & 15.60 & 15.63 & 0.37 & 16.72 & 1.00 & 2.2 & 0.50 & $-$46.6 & 16.12 & 4.0 & 0.41 & $-$80.1 & 1.22 & 1 \\
J01183556$-$0013594 & 0.0455 & S,J & 14.96 & 14.92 & 0.01 & 19.96 & 8.00 & 0.0 & 0.05 & $-$34.5 & 14.93 & 4.3 & 0.91 & 51.3 & \nodata & 1 \\
J02110638$-$0039191 & 0.0177 & S,J & 14.50 & 14.35 & 0.27 & 15.77 & 8.00 & 0.5 & 0.72 & $-$28.3 & 14.69 & 6.9 & 0.16 & $-$70.1 & 1.21 & 1 \\
J02110832$-$0039171 & 0.0181 & S,J & 13.62 & 13.24 & 0.73 & 13.58 & 8.00 & 17.6 & 0.67 & 5.3 & 14.65 & 4.0 & 0.12 & 2.5 & \nodata & 1 \\
J03381222$+$0110088 & 0.0392 & S,I & 15.19 & 15.15 & 0.12 & 17.50 & 5.34 & 1.9 & 0.83 & $-$36.0 & 15.29 & 5.7 & 0.57 & 45.3 & 1.16 & 1 \\
J03381299$+$0109414 & 0.0406 & E,I & 15.82 & 15.71 & 0.82 & 15.93 & 5.48 & 2.1 & 0.60 & 51.3 & 17.58 & 2.1 & 0.76 & 3.5 & 1.15 & 1 \\
J07543194$+$1648214 & 0.0459 & S,M & 14.63 & 14.63 & 0.03 & 18.31 & 1.00 & 0.7 & 0.78 & 28.7 & 14.66 & \nodata & \nodata & \nodata & \nodata & 2 \\
J07543221$+$1648349 & 0.0462 & S,M & 14.53 & 14.53 & 0.15 & 16.60 & 1.21 & 2.1 & 0.43 & $-$7.9 & 14.71 & \nodata & \nodata & \nodata & \nodata & 2 \\
J08083377$+$3854534 & 0.0402 & S,J & 15.45 & 14.99 & 0.86 & 15.15 & 5.97 & 6.1 & 0.83 & $-$10.1 & 17.12 & 6.5 & 0.42 & $-$41.1 & 1.19 & 9 \\
J08083563$+$3854522 & 0.0401 & E,J & 14.61 & 14.53 & 0.60 & 15.08 & 4.56 & 2.8 & 0.90 & 82.1 & 15.53 & 9.3 & 0.73 & 59.0 & \nodata & 1 \\
J08233266$+$2120171 & 0.0181 & S,J & 14.42 & 14.40 & 0.26 & 15.87 & 1.28 & 2.4 & 0.41 & $-$27.5 & 14.72 & 5.0 & 0.73 & 28.5 & 1.21 & 0 \\
J08233421$+$2120515 & 0.0181 & S,J & 13.80 & 13.71 & 0.39 & 14.74 & 8.00 & 7.3 & 0.33 & 62.6 & 14.25 & 9.2 & 0.42 & 53.7 & 1.22 & 0 \\
J08291491$+$5531227 & 0.0251 & S,J & 14.09 & 14.02 & 0.03 & 17.75 & 1.00 & 1.7 & 0.42 & 4.1 & 14.05 & 9.8 & 0.44 & 4.4 & 1.15 & 0 \\
J08292083$+$5531081 & 0.0252 & S,J & 14.04 & 14.07 & 0.11 & 16.50 & 1.77 & 1.4 & 0.38 & 29.4 & 14.19 & 6.1 & 0.52 & 52.0 & 1.16 & 0 \\
J08364482$+$4722188 & 0.0526 & S,J & 15.18 & 15.29 & 0.19 & 17.07 & 1.60 & 0.9 & 0.39 & $-$72.4 & 15.53 & 3.9 & 0.66 & $-$58.3 & 1.14 & 1 \\
J08364588$+$4722100 & 0.0526 & S,J & 14.96 & 14.66 & 0.87 & 14.81 & 5.79 & 6.3 & 0.83 & $-$44.2 & 16.89 & 5.3 & 0.57 & 10.5 & \nodata & 1 \\
J08381759$+$3054534 & 0.0476 & S,I & 15.79 & 15.31 & 0.75 & 15.62 & 8.00 & 10.6 & 0.69 & 21.6 & 16.83 & 3.0 & 0.33 & 58.7 & 1.18 & 9 \\
J08381795$+$3055011 & 0.0481 & S,I & 15.09 & 14.95 & 0.73 & 15.29 & 5.06 & 9.6 & 0.58 & $-$32.7 & 16.39 & 4.1 & 0.27 & $-$43.9 & \nodata & 1 \\
J08385973$+$3613164 & 0.0556 & E,J & 15.07 & 14.74 & 0.78 & 15.01 & 8.00 & 8.4 & 0.86 & 78.6 & 16.36 & 2.0 & 0.39 & 47.5 & 1.13 & 1 \\
J08390125$+$3613042 & 0.0548 & S,J & 15.51 & 15.50 & 0.71 & 15.88 & 2.84 & 4.3 & 0.50 & $-$56.4 & 16.84 & 3.9 & 0.37 & $-$41.3 & \nodata & 1 \\
J08414959$+$2642578 & 0.0848 & S,J & 15.43 & 15.60 & 0.46 & 16.46 & 1.57 & 1.5 & 0.67 & 8.0 & 16.26 & 4.1 & 0.69 & 54.2 & 1.21 & 1 \\
J08415054$+$2642475 & 0.0858 & E,J & 15.71 & 14.87 & 0.91 & 14.97 & 8.00 & 3.2 & 0.64 & $-$47.8 & 17.43 & 2.4 & 0.90 & 26.3 & \nodata & 9 \\
J09060283$+$5144411 & 0.0291 & E,J & 14.48 & 14.41 & 0.59 & 14.99 & 8.00 & 2.4 & 0.25 & 89.8 & 15.38 & 5.2 & 0.34 & $-$82.9 & 1.15 & 0 \\
J09060498$+$5144071 & 0.0291 & S,J & 14.34 & 14.26 & 0.14 & 16.43 & 7.61 & 2.5 & 0.36 & $-$87.3 & 14.42 & 6.7 & 0.88 & 62.0 & 1.16 & 0 \\
J09123636$+$3547180 & 0.0235 & E,J & 14.08 & 13.81 & 0.91 & 13.92 & 5.76 & 7.2 & 0.74 & 52.6 & 16.38 & 30.1 & 0.31 & $-$42.7 & 1.14 & 1 \\
J09123676$+$3547462 & 0.0235 & S,J & 15.04 & 15.04 & 0.37 & 16.12 & 2.42 & 2.0 & 0.62 & 2.4 & 15.55 & 2.8 & 0.96 & 35.4 & 1.20 & 1 \\
J09134461$+$4742165 & 0.0512 & E,I & 14.95 & 14.80 & 0.82 & 15.02 & 3.35 & 3.5 & 0.93 & 41.1 & 16.63 & 6.8 & 0.87 & $-$88.5 & 1.21 & 0 \\
J09134606$+$4742001 & 0.0527 & S,I & 14.89 & 14.80 & 0.43 & 15.71 & 7.52 & 2.5 & 0.46 & 48.1 & 15.41 & 8.0 & 0.43 & 38.4 & 1.24 & 0 \\
J09155467$+$4419510 & 0.0396 & S,M & 14.97 & 14.62 & 1.00 & 14.62 & 1.00 & 12.2 & 0.56 & 82.2 & 99.00 & \nodata & \nodata & \nodata & \nodata & 6 \\
J09155552$+$4419580 & 0.0396 & S,M & 14.14 & 14.14 & 0.20 & 15.91 & 1.00 & 5.3 & 0.39 & 32.6 & 14.38 & \nodata & \nodata & \nodata & \nodata & 2 \\
J09264111$+$0447247 & 0.0891 & S,M & 16.21 & 15.94 & 1.00 & 15.94 & 5.61 & 6.0 & 0.82 & 74.9 & 99.00 & \nodata & \nodata & \nodata & \nodata & 6 \\
J09264137$+$0447260 & 0.0907 & S,M & 16.17 & 16.17 & 0.70 & 16.56 & 2.93 & 2.4 & 0.46 & 51.7 & 17.46 & \nodata & \nodata & \nodata & \nodata & 2 \\
J09374413$+$0245394 & 0.0242 & S,I & 12.85 & 12.85 & 0.09 & 15.44 & 8.00 & 16.8 & 0.24 & 28.4 & 12.96 & \nodata & \nodata & \nodata & \nodata & 2 \\
J09374506$+$0244504 & 0.0235 & E,I & 13.47 & 13.29 & 0.82 & 13.50 & 4.77 & 6.2 & 0.63 & 14.1 & 15.16 & 10.8 & 0.88 & $-$9.0 & 1.21 & 0 \\
J10100079$+$5440198 & 0.0460 & S,M & 14.90 & 14.90 & 0.00 & 22.31 & 3.28 & 0.0 & 0.05 & $-$0.4 & 14.90 & \nodata & \nodata & \nodata & \nodata & 2 \\
J10100212$+$5440279 & 0.0463 & S,M & 15.67 & 15.67 & 0.29 & 17.02 & 2.02 & 1.2 & 0.51 & 42.7 & 16.04 & \nodata & \nodata & \nodata & \nodata & 2 \\
J10155257$+$0657330 & 0.0299 & S,J & 15.09 & 15.68 & 0.33 & 16.89 & 1.71 & 1.3 & 0.42 & $-$42.0 & 16.11 & 2.5 & 0.73 & 25.9 & 1.12 & 9 \\
J10155338$+$0657495 & 0.0291 & E,J & 14.06 & 13.61 & 0.88 & 13.75 & 6.73 & 16.3 & 0.27 & $-$39.2 & 15.94 & 4.4 & 0.78 & 73.3 & \nodata & 1 \\
J10205188$+$4831096 & 0.0530 & S,I & 16.11 & 15.76 & 0.34 & 16.94 & 3.00 & 0.0 & 0.02 & 60.2 & 16.20 & 3.0 & 0.68 & $-$13.9 & 1.18 & 1 \\
J10205369$+$4831246 & 0.0531 & E,I & 15.23 & 14.99 & 0.89 & 15.12 & 2.39 & 7.4 & 0.90 & $-$72.0 & 17.41 & 0.4 & 0.40 & $-$46.7 & \nodata & 1 \\
J10225647$+$3446564 & 0.0554 & S,I & 15.99 & 16.15 & 0.35 & 17.28 & 1.00 & 2.1 & 0.18 & $-$49.2 & 16.63 & 3.1 & 0.52 & $-$64.6 & 1.15 & 1 \\
J10225655$+$3446468 & 0.0564 & S,I & 15.03 & 14.79 & 0.93 & 14.86 & 6.80 & 4.9 & 0.78 & $-$60.8 & 17.73 & 3.3 & 0.50 & $-$67.3 & \nodata & 1 \\
J10233658$+$4220477 & 0.0456 & S,I & 15.00 & 14.93 & 0.14 & 17.08 & 1.62 & 2.0 & 0.25 & $-$84.4 & 15.09 & 4.0 & 0.73 & $-$31.8 & 1.26 & 1 \\
J10233684$+$4221037 & 0.0454 & S,I & 15.84 & 15.92 & 0.37 & 17.01 & 1.00 & 5.7 & 0.76 & $-$58.2 & 16.42 & 0.9 & 0.68 & $-$61.9 & \nodata & 1 \\
J10272950$+$0114490 & 0.0236 & S,I & 14.34 & 14.39 & 0.52 & 15.11 & 1.00 & 4.9 & 0.54 & $-$39.0 & 15.18 & 13.7 & 0.56 & $-$46.0 & 1.21 & 1 \\
J10272970$+$0115170 & 0.0234 & E,I & 13.56 & 13.35 & 0.70 & 13.74 & 4.49 & 7.7 & 0.70 & 55.8 & 14.64 & 14.6 & 0.65 & $-$29.5 & \nodata & 1 \\
J10325316$+$5306536 & 0.0640 & S,I & 15.49 & 15.17 & 0.72 & 15.53 & 8.00 & 34.4 & 0.88 & $-$71.3 & 16.54 & 2.6 & 0.26 & $-$80.1 & 1.18 & 1 \\
J10325321$+$5306477 & 0.0639 & E,I & 15.17 & 14.64 & 0.94 & 14.72 & 6.24 & 9.9 & 0.96 & 65.4 & 17.61 & 1.3 & 0.46 & $-$50.2 & \nodata & 9 \\
J10332972$+$4404342 & 0.0523 & S,J & 14.79 & 14.75 & 0.26 & 16.19 & 1.59 & 2.2 & 0.21 & 72.6 & 15.08 & 3.4 & 0.72 & 76.1 & 1.22 & 0 \\
J10333162$+$4404212 & 0.0521 & S,J & 15.65 & 15.62 & 0.13 & 17.80 & 1.00 & 1.9 & 0.24 & 68.1 & 15.78 & 6.8 & 0.14 & 67.1 & 1.20 & 0 \\
J10364274$+$5447356 & 0.0458 & S,I & 15.34 & 15.34 & 0.79 & 15.59 & 5.57 & 4.2 & 0.82 & 57.1 & 17.04 & \nodata & \nodata & \nodata & \nodata & 2 \\
J10364400$+$5447489 & 0.0458 & E,I & 14.34 & 14.34 & 0.88 & 14.48 & 7.34 & 5.6 & 0.80 & 62.6 & 16.60 & \nodata & \nodata & \nodata & \nodata & 2 \\
J10392338$+$3904501 & 0.0435 & S,J & 15.26 & 15.20 & 0.38 & 16.26 & 1.88 & 0.9 & 0.74 & $-$78.2 & 15.72 & 5.1 & 0.87 & $-$49.4 & 1.14 & 1 \\
J10392515$+$3904573 & 0.0433 & E,J & 14.97 & 14.87 & 0.78 & 15.14 & 4.61 & 3.5 & 0.79 & 12.2 & 16.52 & 7.4 & 0.37 & $-$14.9 & \nodata & 1 \\
J10435053$+$0645466 & 0.0287 & S,I & 14.26 & 14.28 & 0.11 & 16.67 & 2.56 & 0.0 & 0.90 & 55.8 & 14.41 & 4.2 & 1.00 & $-$43.5 & 1.23 & 1 \\
J10435268$+$0645256 & 0.0281 & S,I & 14.66 & 14.73 & 0.23 & 16.35 & 1.96 & 2.9 & 0.69 & 11.9 & 15.01 & 7.0 & 0.67 & 16.2 & \nodata & 1 \\
J10452478$+$3910298 & 0.0268 & S,J & 14.44 & 14.45 & 0.46 & 15.28 & 2.83 & 2.8 & 0.74 & $-$37.1 & 15.12 & 7.1 & 0.35 & $-$10.8 & 1.17 & 1 \\
J10452496$+$3909499 & 0.0257 & E,J & 13.91 & 13.59 & 0.75 & 13.90 & 8.00 & 13.2 & 0.89 & 14.2 & 15.12 & 13.1 & 0.57 & 69.1 & \nodata & 1 \\
J10514368$+$5101195 & 0.0250 & E,J & 12.94 & 12.72 & 0.66 & 13.17 & 6.04 & 11.7 & 0.69 & 74.5 & 13.90 & 27.5 & 0.49 & $-$69.1 & 1.08 & 1 \\
J10514450$+$5101303 & 0.0238 & S,J & 14.06 & 13.95 & 0.56 & 14.58 & 3.09 & 6.3 & 0.53 & 22.5 & 14.84 & 10.5 & 0.50 & $-$9.5 & \nodata & 1 \\
J10595869$+$0857215 & 0.0632 & E,J & 15.02 & 14.67 & 0.81 & 14.90 & 7.05 & 6.9 & 0.95 & 61.3 & 16.50 & 8.1 & 0.55 & $-$42.9 & 1.17 & 1 \\
J10595915$+$0857357 & 0.0627 & S,J & 15.63 & 15.61 & 0.48 & 16.40 & 8.00 & 2.8 & 0.52 & 28.8 & 16.32 & 3.6 & 0.27 & 35.4 & \nodata & 1 \\
J11014357$+$5720058 & 0.0469 & E,J & 15.14 & 14.94 & 0.84 & 15.13 & 8.00 & 3.5 & 0.78 & $-$85.7 & 16.94 & 3.1 & 0.58 & $-$14.3 & 1.15 & 1 \\
J11014364$+$5720336 & 0.0478 & S,J & 15.85 & 15.83 & 0.55 & 16.49 & 2.66 & 0.9 & 0.76 & 48.7 & 16.69 & 4.0 & 0.82 & $-$20.5 & \nodata & 1 \\
J11064944$+$4751119 & 0.0643 & S,I & 15.48 & 15.35 & 0.87 & 15.50 & 5.88 & 1.7 & 0.78 & 29.1 & 17.58 & 2.2 & 0.35 & 30.8 & 1.20 & 1 \\
J11065068$+$4751090 & 0.0654 & S,I & 15.23 & 15.29 & 0.87 & 15.44 & 1.00 & 6.7 & 0.45 & 4.7 & 17.51 & 0.5 & 0.63 & $-$21.1 & \nodata & 1 \\
J11204657$+$0028142 & 0.0255 & S,J & 14.37 & 14.10 & 0.96 & 14.14 & 8.00 & 4.6 & 0.62 & 11.2 & 17.62 & 1.8 & 0.18 & 24.8 & 1.17 & 1 \\
J11204801$+$0028068 & 0.0256 & S,J & 13.86 & 13.95 & 0.47 & 14.78 & 3.42 & 2.1 & 0.62 & $-$34.7 & 14.63 & 8.7 & 0.15 & $-$20.2 & \nodata & 1 \\
J11251704$+$0227007 & 0.0504 & S,I & 15.73 & 15.85 & 0.52 & 16.57 & 5.15 & 0.6 & 0.35 & $-$60.7 & 16.64 & 2.6 & 0.42 & $-$28.7 & 1.18 & 1 \\
J11251716$+$0226488 & 0.0507 & S,I & 15.01 & 14.92 & 0.59 & 15.49 & 2.39 & 2.7 & 0.64 & $-$79.8 & 15.90 & 7.4 & 0.55 & 32.3 & \nodata & 1 \\
J11273289$+$3604168 & 0.0351 & S,J & 14.36 & 14.28 & 0.54 & 14.95 & 4.59 & 2.8 & 0.49 & $-$6.4 & 15.12 & 7.5 & 0.83 & 68.0 & 1.15 & 0 \\
J11273467$+$3603470 & 0.0351 & S,J & 13.86 & 13.80 & 0.11 & 16.19 & 3.60 & 1.3 & 0.45 & 12.9 & 13.93 & 8.3 & 0.30 & 56.6 & 1.20 & 0 \\
J11375476$+$4727588 & 0.0343 & E,J & 14.08 & 13.87 & 0.86 & 14.04 & 4.43 & 4.9 & 0.74 & 57.5 & 15.99 & 21.7 & 0.40 & 73.1 & 1.17 & 1 \\
J11375801$+$4728143 & 0.0339 & S,J & 14.64 & 14.74 & 0.59 & 15.32 & 2.52 & 1.7 & 0.60 & 18.9 & 15.70 & 5.9 & 0.17 & 20.3 & \nodata & 1 \\
J11440335$+$3332062 & 0.0318 & E,I & 14.73 & 14.62 & 0.58 & 15.22 & 6.69 & 5.0 & 0.51 & $-$1.3 & 15.56 & 3.4 & 0.50 & $-$20.2 & 1.16 & 0 \\
J11440433$+$3332339 & 0.0315 & S,I & 15.38 & 15.42 & 0.15 & 17.46 & 1.36 & 1.3 & 0.43 & 81.6 & 15.60 & 3.0 & 0.55 & $-$14.1 & 1.17 & 0 \\
J11484370$+$3547002 & 0.0641 & S,I & 16.50 & 16.46 & 0.23 & 18.05 & 1.25 & 3.6 & 0.50 & 75.9 & 16.74 & 6.1 & 0.16 & 82.2 & 1.22 & 1 \\
J11484525$+$3547092 & 0.0636 & S,I & 14.77 & 14.73 & 0.13 & 16.93 & 2.85 & 2.9 & 0.28 & $-$37.4 & 14.88 & 4.9 & 0.79 & $-$12.8 & \nodata & 1 \\
J11501333$+$3746107 & 0.0550 & S,I & 15.56 & 15.53 & 0.54 & 16.20 & 2.18 & 1.8 & 0.65 & $-$63.4 & 16.37 & 4.3 & 0.55 & 54.1 & 1.17 & 1 \\
J11501399$+$3746306 & 0.0555 & S,I & 15.19 & 15.18 & 0.10 & 17.71 & 4.87 & 0.8 & 0.05 & 0.6 & 15.29 & 2.4 & 0.75 & 23.6 & \nodata & 1 \\
J11505764$+$1444200 & 0.0572 & S,I & 15.94 & 15.83 & 0.59 & 16.41 & 2.42 & 2.0 & 0.61 & $-$83.9 & 16.80 & 8.4 & 0.82 & $-$81.9 & 1.16 & 1 \\
J11505844$+$1444124 & 0.0562 & E,I & 14.53 & 14.32 & 0.90 & 14.43 & 6.07 & 5.4 & 0.83 & $-$85.3 & 16.84 & 7.6 & 0.55 & 52.0 & \nodata & 1 \\
J11542299$+$4932509 & 0.0702 & S,I & 15.92 & 15.69 & 0.21 & 17.38 & 8.00 & 0.8 & 0.48 & $-$5.1 & 15.95 & 18.4 & 0.34 & $-$23.9 & 1.14 & 1 \\
J11542307$+$4932456 & 0.0712 & E,I & 15.35 & 15.12 & 0.59 & 15.70 & 8.00 & 2.3 & 0.55 & $-$23.4 & 16.08 & 7.8 & 0.86 & 27.9 & \nodata & 1 \\
J12020424$+$5342317 & 0.0647 & S,I & 16.01 & 15.90 & 0.89 & 16.03 & 3.80 & 6.3 & 0.61 & 33.1 & 18.27 & 2.3 & 0.18 & 41.2 & 1.20 & 1 \\
J12020537$+$5342487 & 0.0640 & E,I & 15.21 & 15.03 & 0.83 & 15.22 & 8.00 & 5.7 & 0.71 & $-$56.6 & 16.97 & 3.9 & 0.77 & $-$32.8 & \nodata & 1 \\
J12054066$+$0135365 & 0.0220 & S,J & 14.52 & 14.49 & 0.26 & 15.98 & 1.60 & 1.7 & 0.81 & 20.0 & 14.81 & 4.5 & 0.77 & $-$87.5 & 1.15 & 0 \\
J12054073$+$0134302 & 0.0209 & E,J & 14.05 & 14.01 & 0.51 & 14.75 & 2.22 & 3.2 & 0.71 & $-$76.8 & 14.77 & 10.2 & 0.22 & $-$83.5 & 1.16 & 0 \\
J12115507$+$4039182 & 0.0229 & S,I & 14.51 & 14.35 & 0.40 & 15.35 & 8.00 & 9.2 & 0.82 & 42.9 & 14.90 & 2.6 & 0.46 & 53.2 & 1.21 & 1 \\
J12115648$+$4039184 & 0.0235 & S,I & 14.87 & 14.89 & 0.13 & 17.08 & 8.00 & 4.3 & 0.79 & 53.3 & 15.04 & 3.0 & 0.49 & $-$9.0 & \nodata & 1 \\
J12191719$+$1200582 & 0.0273 & E,I & 14.61 & 14.43 & 0.56 & 15.06 & 5.65 & 2.7 & 0.67 & 82.5 & 15.33 & 7.6 & 0.98 & 35.1 & 1.14 & 1 \\
J12191866$+$1201054 & 0.0268 & S,I & 15.14 & 15.24 & 0.37 & 16.32 & 1.00 & 3.3 & 0.27 & 48.8 & 15.73 & 5.8 & 0.45 & 84.3 & \nodata & 1 \\
J12433887$+$4405399 & 0.0418 & S,J & 14.84 & 14.84 & 0.15 & 16.88 & 4.02 & 2.0 & 0.27 & 62.6 & 15.03 & 5.1 & 0.71 & $-$39.2 & 1.15 & 1 \\
J12433936$+$4406046 & 0.0410 & E,J & 14.49 & 14.17 & 0.90 & 14.28 & 7.26 & 11.3 & 0.57 & 83.3 & 16.68 & 4.3 & 0.77 & $-$62.0 & \nodata & 1 \\
J12525011$+$4645272 & 0.0613 & S,J & 15.47 & 15.53 & 0.10 & 18.05 & 1.00 & 0.4 & 0.88 & $-$88.5 & 15.65 & 2.4 & 0.90 & 14.7 & 1.18 & 1 \\
J12525212$+$4645294 & 0.0610 & E,J & 15.11 & 14.97 & 0.77 & 15.25 & 4.13 & 4.0 & 0.92 & $-$12.9 & 16.57 & 5.6 & 0.72 & 60.3 & \nodata & 1 \\
J13011662$+$4803366 & 0.0303 & S,I & 14.55 & 14.55 & 0.50 & 15.31 & 1.00 & 9.6 & 0.66 & 19.6 & 15.29 & \nodata & \nodata & \nodata & \nodata & 2 \\
J13011835$+$4803304 & 0.0298 & S,I & 15.16 & 15.16 & 0.22 & 16.78 & 1.00 & 11.5 & 0.50 & 51.4 & 15.43 & \nodata & \nodata & \nodata & \nodata & 2 \\
J13082737$+$0422125 & 0.0255 & S,J & 15.61 & 15.59 & 0.03 & 19.35 & 4.82 & 0.0 & 0.30 & $-$52.1 & 15.62 & 4.1 & 0.35 & 80.7 & 1.16 & 1 \\
J13082964$+$0422045 & 0.0257 & S,J & 14.80 & 14.70 & 0.40 & 15.70 & 8.00 & 1.7 & 0.68 & $-$26.0 & 15.26 & 5.4 & 0.70 & 69.1 & \nodata & 1 \\
J13131429$+$3910360 & 0.0716 & E,I & 15.37 & 15.33 & 0.36 & 16.43 & 2.08 & 0.9 & 0.87 & 41.2 & 15.81 & 3.4 & 0.96 & 84.1 & 1.14 & 1 \\
J13131470$+$3910382 & 0.0716 & S,I & 16.36 & 15.90 & 0.43 & 16.82 & 2.78 & 1.0 & 0.47 & 25.7 & 16.50 & 16.0 & 0.93 & 80.7 & \nodata & 9 \\
J13151386$+$4424264 & 0.0359 & S,I & 14.55 & 14.61 & 0.20 & 16.35 & 3.01 & 1.3 & 0.50 & $-$10.3 & 14.86 & 4.6 & 0.72 & 60.6 & 1.22 & 1 \\
J13151726$+$4424255 & 0.0357 & S,I & 14.26 & 14.08 & 0.94 & 14.15 & 5.59 & 8.2 & 0.56 & 81.1 & 17.13 & 9.8 & 0.12 & 66.1 & \nodata & 1 \\
J13153076$+$6207447 & 0.0306 & S,I & 14.59 & 14.59 & 0.26 & 16.04 & 1.00 & 2.3 & 0.51 & $-$75.5 & 14.92 & \nodata & \nodata & \nodata & \nodata & 2 \\
J13153506$+$6207287 & 0.0306 & S,I & 14.57 & 14.57 & 0.16 & 16.56 & 1.00 & 0.9 & 0.09 & $-$62.1 & 14.76 & \nodata & \nodata & \nodata & \nodata & 2 \\
J13325525$-$0301347 & 0.0493 & S,I & 15.61 & 15.58 & 0.51 & 16.30 & 1.00 & 3.7 & 0.29 & 87.1 & 16.37 & 4.9 & 0.42 & $-$76.7 & 1.19 & 1 \\
J13325655$-$0301395 & 0.0483 & S,I & 14.81 & 14.76 & 0.05 & 17.95 & 8.00 & 1.4 & 0.11 & 77.4 & 14.82 & 4.0 & 0.89 & 21.2 & \nodata & 1 \\
J13462001$-$0325407 & 0.0248 & S,J & 13.97 & 13.93 & 0.68 & 14.34 & 3.95 & 7.7 & 0.40 & 7.4 & 15.17 & 6.9 & 0.42 & $-$5.0 & 1.15 & 1 \\
J13462215$-$0325057 & 0.0255 & E,J & 14.17 & 13.84 & 0.89 & 13.97 & 5.66 & 18.3 & 0.62 & 0.7 & 16.19 & 0.8 & 0.19 & $-$3.5 & \nodata & 1 \\
J14003661$-$0254327 & 0.0256 & S,I & 14.26 & 14.21 & 0.70 & 14.59 & 4.76 & 8.5 & 0.72 & $-$44.1 & 15.51 & 4.7 & 0.31 & $-$71.2 & 1.21 & 1 \\
J14003796$-$0254227 & 0.0269 & S,I & 14.76 & 14.55 & 0.49 & 15.33 & 3.76 & 1.9 & 0.58 & 13.0 & 15.28 & 7.1 & 0.60 & $-$1.4 & \nodata & 1 \\
J14005783$+$4251203 & 0.0327 & S,I & 15.10 & 15.10 & 0.02 & 19.26 & 1.00 & 0.2 & 0.84 & 49.0 & 15.12 & 5.4 & 0.29 & $-$11.3 & 1.20 & 1 \\
J14005879$+$4250427 & 0.0335 & S,I & 15.25 & 15.25 & 0.62 & 15.78 & 1.00 & 6.2 & 0.97 & 22.7 & 16.29 & 1.7 & 0.34 & 69.2 & \nodata & 1 \\
J14055079$+$6542598 & 0.0306 & S,J & 15.49 & 15.50 & 0.29 & 16.85 & 1.36 & 1.6 & 0.66 & 61.9 & 15.87 & 5.3 & 0.28 & 53.3 & 1.16 & 0 \\
J14055334$+$6542277 & 0.0308 & E,J & 14.55 & 14.17 & 0.89 & 14.30 & 7.58 & 11.6 & 0.67 & $-$58.3 & 16.57 & 2.3 & 0.67 & 78.5 & 1.09 & 0 \\
J14062157$+$5043303 & 0.0065 & S,J & 12.00 & 11.96 & 0.05 & 15.20 & 1.09 & 1.5 & 0.80 & 78.9 & 12.02 & 15.6 & 0.74 & 9.7 & 1.85 & 0 \\
J14064127$+$5043239 & 0.0073 & E,J & 12.12 & 11.98 & 0.58 & 12.57 & 8.00 & 12.8 & 0.90 & 85.5 & 12.93 & 17.1 & 0.65 & $-$68.8 & 1.11 & 0 \\
J14070703$-$0234513 & 0.0586 & S,I & 15.55 & 15.70 & 0.13 & 17.90 & 3.40 & 0.0 & 0.19 & $-$81.4 & 15.85 & 4.4 & 0.30 & 81.7 & 1.15 & 1 \\
J14070720$-$0234402 & 0.0576 & E,I & 16.02 & 15.60 & 0.75 & 15.91 & 8.00 & 2.4 & 0.78 & 36.2 & 17.09 & 11.0 & 0.37 & 24.4 & \nodata & 1 \\
J14234238$+$3400324 & 0.0136 & S,J & 13.55 & 13.53 & 0.28 & 14.93 & 1.42 & 4.1 & 0.53 & $-$80.2 & 13.88 & 7.5 & 0.87 & 3.7 & 1.40 & 0 \\
J14234632$+$3401012 & 0.0126 & S,J & 13.91 & 13.87 & 0.17 & 15.80 & 1.74 & 2.7 & 0.60 & 10.8 & 14.07 & 8.1 & 0.30 & 35.9 & 1.28 & 0 \\
J14245831$-$0303597 & 0.0525 & S,J & 15.02 & 15.03 & 0.05 & 18.32 & 3.25 & 0.1 & 0.08 & 51.0 & 15.09 & 2.6 & 0.77 & 41.2 & 1.18 & 1 \\
J14245913$-$0304012 & 0.0535 & S,J & 15.00 & 14.90 & 0.18 & 16.75 & 2.05 & 1.7 & 0.34 & 33.8 & 15.12 & 10.9 & 0.20 & 37.0 & \nodata & 1 \\
J14250552$+$0313590 & 0.0371 & E,I & 14.53 & 14.22 & 1.00 & 14.22 & 6.55 & 17.9 & 0.78 & 78.3 & 99.00 & \nodata & \nodata & \nodata & \nodata & 6 \\
J14250739$+$0313560 & 0.0375 & S,I & 15.64 & 15.64 & 0.55 & 16.29 & 5.90 & 0.9 & 0.74 & $-$4.7 & 16.50 & \nodata & \nodata & \nodata & \nodata & 2 \\
J14294766$+$3534275 & 0.0290 & S,J & 13.75 & 13.70 & 0.56 & 14.32 & 6.00 & 5.4 & 0.55 & 74.4 & 14.59 & 7.0 & 0.19 & 67.8 & 1.20 & 0 \\
J14295031$+$3534122 & 0.0296 & S,J & 14.62 & 14.66 & 0.14 & 16.81 & 1.02 & 2.3 & 0.56 & $-$61.6 & 14.82 & 5.0 & 0.42 & 12.7 & 1.22 & 0 \\
J14334683$+$4004512 & 0.0260 & S,I & 13.48 & 13.30 & 0.44 & 14.20 & 8.00 & 32.6 & 0.64 & $-$31.7 & 13.93 & 8.7 & 0.46 & $-$64.0 & 1.24 & 1 \\
J14334840$+$4005392 & 0.0264 & S,I & 13.91 & 13.79 & 0.80 & 14.04 & 4.01 & 4.6 & 0.68 & 27.2 & 15.54 & 12.3 & 0.78 & 88.5 & \nodata & 1 \\
J14442055$+$1207429 & 0.0304 & S,M & 14.44 & 14.44 & 0.42 & 15.38 & 6.00 & 5.4 & 0.57 & $-$47.0 & 15.03 & \nodata & \nodata & \nodata & \nodata & 2 \\
J14442079$+$1207552 & 0.0314 & S,M & 13.59 & 13.59 & 0.28 & 14.96 & 6.08 & 7.0 & 0.75 & $-$29.9 & 13.95 & \nodata & \nodata & \nodata & \nodata & 2 \\
J15002374$+$4316559 & 0.0311 & E,J & 14.08 & 13.76 & 0.69 & 14.16 & 5.04 & 6.1 & 0.91 & 88.4 & 15.03 & 16.3 & 0.73 & $-$7.0 & 1.17 & 1 \\
J15002500$+$4317131 & 0.0316 & S,J & 14.57 & 14.77 & 0.40 & 15.75 & 2.02 & 1.5 & 0.80 & $-$29.5 & 15.33 & 3.1 & 0.63 & $-$60.9 & \nodata & 1 \\
J15053137$+$3427534 & 0.0745 & S,I & 15.80 & 15.69 & 0.84 & 15.88 & 8.00 & 2.0 & 0.44 & $-$89.8 & 17.67 & 4.0 & 0.22 & $-$87.5 & 1.17 & 1 \\
J15053183$+$3427526 & 0.0735 & E,I & 15.44 & 15.21 & 1.00 & 15.21 & 6.16 & 4.6 & 0.61 & $-$50.9 & 28.88 & 0.5 & 0.05 & $-$62.5 & \nodata & 1 \\
J15064391$+$0346364 & 0.0363 & S,J & 14.28 & 14.20 & 0.54 & 14.87 & 8.00 & 1.9 & 0.34 & 81.6 & 15.04 & 6.0 & 0.50 & 71.6 & 1.19 & 0 \\
J15064579$+$0346214 & 0.0352 & S,J & 14.68 & 14.60 & 0.46 & 15.44 & 4.20 & 4.4 & 0.34 & $-$79.6 & 15.27 & 12.7 & 0.16 & $-$86.8 & 1.19 & 0 \\
J15101587$+$5810425 & 0.0303 & S,J & 14.63 & 14.46 & 0.35 & 15.59 & 2.29 & 1.2 & 0.72 & $-$36.7 & 14.93 & 4.7 & 0.91 & 73.2 & 1.20 & 0 \\
J15101776$+$5810375 & 0.0317 & S,J & 15.65 & 15.56 & 0.71 & 15.93 & 1.64 & 8.7 & 0.27 & $-$28.4 & 16.89 & 2.6 & 0.21 & $-$35.4 & 1.21 & 0 \\
J15144544$+$0403587 & 0.0386 & S,I & 15.04 & 15.03 & 0.52 & 15.73 & 4.43 & 2.2 & 0.74 & 12.5 & 15.84 & 3.9 & 0.15 & 7.0 & 1.19 & 1 \\
J15144697$+$0403576 & 0.0392 & S,I & 14.82 & 14.73 & 0.48 & 15.52 & 2.96 & 1.9 & 0.75 & 76.4 & 15.45 & 8.6 & 0.42 & $-$84.1 & \nodata & 1 \\
J15233768$+$3749030 & 0.0234 & S,I & 15.04 & 15.12 & 0.16 & 17.08 & 1.00 & 4.4 & 0.22 & 65.7 & 15.32 & 3.7 & 0.54 & $-$75.8 & 1.16 & 1 \\
J15233899$+$3748254 & 0.0236 & E,I & 14.86 & 14.66 & 0.74 & 15.00 & 2.96 & 12.3 & 0.57 & $-$31.2 & 16.11 & 2.0 & 0.78 & $-$24.1 & \nodata & 1 \\
J15264774$+$5915464 & 0.0447 & S,I & 15.07 & 14.86 & 0.81 & 15.09 & 8.00 & 5.2 & 0.62 & $-$68.3 & 16.64 & 7.9 & 0.24 & 73.9 & 1.14 & 1 \\
J15264892$+$5915478 & 0.0455 & E,I & 14.96 & 14.90 & 0.85 & 15.08 & 8.00 & 4.8 & 0.58 & $-$80.7 & 16.95 & 3.8 & 0.93 & 14.9 & \nodata & 1 \\
J15281276$+$4255474 & 0.0188 & S,I & 13.12 & 13.20 & 0.31 & 14.47 & 1.00 & 9.1 & 0.19 & $-$0.6 & 13.61 & 8.3 & 0.45 & 5.1 & 1.45 & 1 \\
J15281667$+$4256384 & 0.0180 & S,I & 13.40 & 13.19 & 0.77 & 13.47 & 8.00 & 12.6 & 0.56 & 33.3 & 14.80 & 4.0 & 0.19 & 26.9 & \nodata & 1 \\
J15523258$+$4620180 & 0.0594 & E,I & 15.10 & 14.85 & 0.76 & 15.15 & 7.31 & 5.3 & 0.71 & 1.2 & 16.38 & 5.9 & 0.84 & $-$59.7 & 1.18 & 1 \\
J15523393$+$4620237 & 0.0610 & S,I & 15.60 & 15.64 & 0.27 & 17.08 & 8.00 & 4.9 & 0.44 & 65.4 & 15.97 & 3.5 & 0.40 & 80.5 & \nodata & 1 \\
J15562191$+$4757172 & 0.0191 & S,J & 14.71 & 14.65 & 0.04 & 18.26 & 1.00 & 0.4 & 0.31 & $-$70.8 & 14.69 & 7.0 & 0.45 & $-$70.2 & 1.18 & 0 \\
J15562738$+$4757302 & 0.0199 & E,J & 14.87 & 14.73 & 0.51 & 15.47 & 8.00 & 4.6 & 0.62 & $-$73.6 & 15.51 & 2.3 & 0.75 & $-$51.2 & 1.14 & 0 \\
J15583749$+$3227379 & 0.0494 & S,I & 16.15 & 16.23 & 0.47 & 17.06 & 1.00 & 1.3 & 0.78 & 21.3 & 16.91 & 3.2 & 0.26 & 29.8 & 1.16 & 1 \\
J15583784$+$3227471 & 0.0485 & S,I & 15.18 & 15.05 & 0.60 & 15.60 & 1.05 & 4.3 & 0.45 & $-$17.9 & 16.05 & 8.4 & 0.50 & $-$7.0 & \nodata & 1 \\
J16024254$+$4111499 & 0.0335 & S,J & 14.14 & 14.11 & 0.08 & 16.85 & 3.82 & 3.8 & 0.21 & 50.0 & 14.20 & 6.0 & 0.63 & 18.5 & 1.34 & 0 \\
J16024475$+$4111589 & 0.0333 & S,J & 15.18 & 15.08 & 0.02 & 19.12 & 1.00 & 1.0 & 0.23 & $-$7.0 & 15.11 & 4.5 & 0.37 & 68.1 & 1.22 & 0 \\
J16080559$+$2529091 & 0.0415 & S,M & 15.02 & 15.02 & 0.15 & 17.12 & 1.00 & 0.4 & 0.62 & $-$86.3 & 15.19 & \nodata & \nodata & \nodata & \nodata & 2 \\
J16080648$+$2529066 & 0.0423 & S,M & 14.71 & 14.71 & 0.39 & 15.73 & 1.00 & 5.3 & 0.74 & $-$45.5 & 15.24 & \nodata & \nodata & \nodata & \nodata & 2 \\
J16082261$+$2328459 & 0.0409 & S,J & 15.00 & 14.88 & 0.04 & 18.45 & 8.00 & 2.4 & 0.23 & $-$42.4 & 14.93 & 7.4 & 0.78 & 84.7 & 1.19 & 1 \\
J16082354$+$2328240 & 0.0408 & S,J & 15.50 & 15.52 & 0.25 & 17.03 & 6.36 & 2.9 & 0.46 & $-$54.2 & 15.83 & 2.6 & 0.68 & 54.5 & \nodata & 1 \\
J16145418$+$3711064 & 0.0582 & S,I & 14.64 & 14.62 & 0.74 & 14.94 & 5.08 & 5.6 & 0.73 & $-$62.4 & 16.10 & 0.0 & 0.21 & $-$57.6 & 1.24 & 1 \\
J16145421$+$3711136 & 0.0582 & E,I & 15.18 & 14.87 & 0.46 & 15.73 & 3.76 & 2.2 & 0.87 & 15.0 & 15.53 & 7.4 & 0.99 & $-$75.0 & \nodata & 1 \\
J16282497$+$4110064 & 0.0330 & S,J & 14.01 & 13.85 & 0.41 & 14.82 & 2.91 & 6.6 & 0.48 & 56.8 & 14.43 & 10.1 & 0.96 & $-$9.3 & 1.16 & 1 \\
J16282756$+$4109395 & 0.0318 & S,J & 14.19 & 14.02 & 0.55 & 14.67 & 8.00 & 16.6 & 0.82 & $-$57.8 & 14.89 & 3.2 & 0.65 & $-$88.1 & \nodata & 1 \\
J16354293$+$2630494 & 0.0701 & S,I & 15.22 & 14.91 & 0.66 & 15.35 & 8.00 & 10.5 & 0.72 & 33.5 & 16.09 & 4.8 & 0.46 & $-$41.5 & 1.18 & 1 \\
J16354366$+$2630505 & 0.0713 & E,I & 15.57 & 15.44 & 0.49 & 16.22 & 2.62 & 1.4 & 0.68 & $-$1.6 & 16.16 & 9.1 & 0.51 & $-$89.8 & \nodata & 1 \\
J16372583$+$4650161 & 0.0578 & S,J & 14.99 & 14.68 & 0.49 & 15.46 & 8.00 & 14.5 & 0.98 & $-$21.2 & 15.42 & 4.7 & 0.31 & 0.1 & 1.16 & 1 \\
J16372754$+$4650054 & 0.0568 & S,J & 15.14 & 15.16 & 0.02 & 19.17 & 1.00 & 0.6 & 0.75 & $-$12.3 & 15.18 & 7.8 & 0.30 & $-$1.3 & \nodata & 1 \\
J17020320$+$1900006 & 0.0573 & E,I & 15.09 & 14.79 & 0.38 & 15.83 & 4.37 & 1.7 & 0.84 & 42.7 & 15.32 & 10.7 & 0.66 & $-$53.6 & 1.15 & 1 \\
J17020378$+$1859495 & 0.0558 & S,I & 16.55 & 16.69 & 0.36 & 17.81 & 8.00 & 1.4 & 0.41 & $-$37.4 & 17.17 & 4.5 & 0.73 & 64.4 & \nodata & 1 \\
J17045089$+$3448530 & 0.0572 & S,I & 15.99 & 15.56 & 0.48 & 16.34 & 4.39 & 2.6 & 0.56 & $-$42.1 & 16.27 & 8.6 & 0.45 & 30.1 & 1.16 & 1 \\
J17045097$+$3449020 & 0.0563 & S,I & 15.26 & 15.45 & 0.28 & 16.85 & 1.00 & 0.5 & 0.81 & 58.0 & 15.81 & 2.2 & 0.88 & $-$87.2 & \nodata & 1 \\
J20471908$+$0019150 & 0.0130 & S,J & 11.85 & 11.47 & 0.34 & 12.66 & 3.31 & 8.4 & 0.82 & 40.9 & 11.92 & 38.5 & 0.64 & 70.2 & 1.28 & 1 \\
J20472428$+$0018030 & 0.0116 & E,J & 12.16 & 12.46 & 0.68 & 12.89 & 4.82 & 12.4 & 0.75 & $-$15.1 & 13.68 & 10.9 & 0.60 & $-$13.8 & \nodata & 9 \\
\enddata
\tablecomments{The columns are: (1) galaxy name; (2) redshift (after the correction for Virgocentric flow); (3) galaxy morphology: 'S' for Spiral, 'E' for Elliptical, and interaction type \citep{2016ApJS..222...16C}: 'J' for JUS, 'I' for INT, 'M' for MER; (4) magnitude taken from \textsc{se}xtractor $\rm 25\;mag\;arcsec^{-2}$ isophot or polygon aperture;} (5) model magnitude taken from \textsc{galfit}, except for distorted galaxies (with flag bit 2) for which this is exactly the same as observed magnitude; (6) Bulge$-$to$-$Total ratio (B/T); (7) bulge magnitude; (8) bulge S\'ersic index; (9) bulge effective radius, noting that all the radii (re or rs) less than 0.05 are recorded as 0.0 in this table due to the limit of decimal places; (10) bulge axis ratio; (11) bulge position angle (the counterclockwise angle between the major axis of the ellipse and North, and the same for disk position angle); (12) disk magnitude; (13) disk scale length; (14) disk axis ratio; (15) disk position angle; (16) reduced Chi$-$square; (17) binary flag in decimal number: Bit 0 (0, 0x0) — galaxy fitted separately, Bit 1 (1, 0x1) — galaxy fitted in pair, Bit 2 (2, 0x10) — distorted galaxy with the bulge fitted by model and the total flux measured by photometry, Bit 3 (4, 0x100) — the fitted bulge is brighter than the total flux of the galaxy obtained in aperture photometry, Bit 4 (8, 0x1000) — the magnitude difference between model and isophot is significant.
\end{deluxetable*}

\begin{figure}[htb!]
\includegraphics[width=0.47\textwidth]{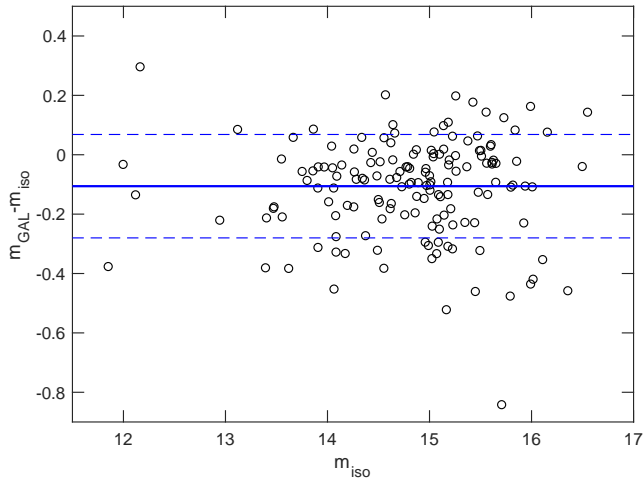}
\caption{Magnitude differences between \textsc{galfit} model and \textsc{se}xtractor $\rm 25\;mag\;arcsec^{-2}$ isophotal photometry for paired galaxies. The solid line represents the mean of the differences. The dashed lines are 1$\sigma$ of the scatters.
\label{fig:mod-obs}}
\end{figure}

As a quality check, in Figure~\ref{fig:mod-obs} we compare the \textsc{se}tractor $\rm 25\;mag\;arcsec^{-2}$ isophotal photometry magnitude and model magnitude (the sum of the bulges and the disks) of 155 galaxies that are fitted using standard \textsc{galfit}. Because the isophotal photometry may miss fluxes from the outskirts of galaxies, especially of ellipticals, while the \textsc{galfit} models are extended to infinity, there is an offset between the isophotal and model magnitudes. For paired galaxies in our sample, the \textsc{galfit} magnitude $\rm m_{GAL}$ are on average brighter than the isophotal magnitudes by $-$0.11 mag with a standard deviation of 0.17 mag. There are 7 galaxies with large $\rm mag_{GAL}- mag_{iso}$ deviations from the mean ($> 2 \sigma$) and they are flagged with it Bit 4 (8, 0x1000) in Table \ref{tab:catalog}. The only one with $> 3 \sigma$ deviation is contaminated from a nearby galaxy. We note that the \textsc{galfit} results of these galaxies should be taken with caution.

Bulges can be separated into pseudo-bulges and classical bulges. The pseudo-bulges have S\'ersic index n = 1–2 \citep*{2004ARA&A..42..603K}. While the classical bulges may stabilize the gas disk and suppress the star formation in paired galaxies \citep{*1996ApJ...464..641M} and in normal galaxies \citep{2009ApJ...707..250M}, the pseudo-bulges are commonly associated with bars and nuclear disks or rings \citep*{2004ARA&A..42..603K} which are disk phenomena, and which themselves may be triggered by interaction and associated with enhanced nuclear star formation \citep{2018ApJS..239...10C,2021MNRAS.502.2446E}. Therefore, we treat all galaxies in our pair sample with bulge S\'ersic index $\rm n \leq 2$ as disk-only and assign $\rm B/T=0$ to them hereinafter. It is worth noting that some pseudo-bulges are not related to nuclear star formation and are mostly found in late-type spirals with relatively low B/T ratios \citep{2016ApJS..225....6K}. Treating them as disky galaxies (i.e. $\rm B/T=0$) shall not introduce significant bias in our results.

\begin{figure}[htb!]
\includegraphics[width=0.47\textwidth]{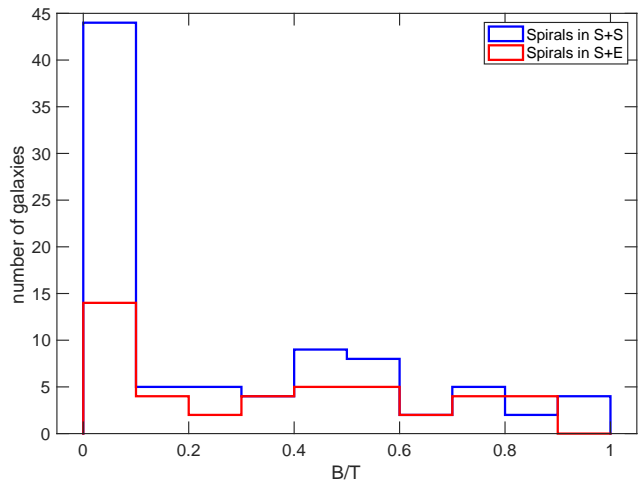}
\caption{Histogram of B/T of spiral galaxies in two kinds of pairs.
\label{fig:bt_hist}}
\end{figure}

For the sSFR enhancement analysis, we are only interested in the B/T ratios of spiral galaxies. Figure~\ref{fig:bt_hist} presents the histograms of B/T ratio distributions of spiral galaxies in S+S and S+E pairs separately. We find spiral galaxies in S+E pairs have larger bulges (at a mean $\rm B/T = 0.35 \pm 0.05$) than their counterparts in S+S pairs (at a mean $\rm B/T = 0.26 \pm 0.03$) statistically. In particular, $50\%$ (44/88) of spirals in S+S pairs are disky galaxies with $\rm B/T<0.1$, while the number in S+E pairs is $32\%$ (14/44). On the other hand, $34\%$ (15/44) of spirals in S+E pairs are bulge dominated with $\rm B/T>0.5$, while this percentage for those in S+S pairs is $24\%$ (21/88).

\section{Control Sample} \label{sec:ctrl}
In order to quantify the sSFR enhancement, a control sample of single spiral galaxies is selected from the catalog of \citet{2015MNRAS.446.3943M}. For each paired spiral galaxy 10 control galaxies are selected. Every control galaxy must meet the following criteria:
\begin{enumerate}
\item Should be identified as spiral in Galaxy Zoo \citep{2008MNRAS.389.1179L}.
\item Not in any interacting system, namely no neighbor galaxy in the SDSS database which has projected distance $\rm \leq 100\;kpc$ and observed redshift difference $\rm \leq 1000\;km\;s^{-1}$.
\item Has reliable B/T ratio, namely having $\chi^2/\nu < 2$ and no bad flag (flag bit 20 = 0) in \citet{2015MNRAS.446.3943M}.
\item The $\rm L_K$ matches that of the paired galaxy within 0.1~dex (or 0.2~dex for controls of 5 H-KPAIR galaxies which have too few control candidates).
\item The B/T ratio matches that of the paired galaxy with $\rm \delta(B/T) < 0.1$, except for paired galaxies of $\rm B/T \geq 0.8$ whose controls shall also have $\rm B/T \geq 0.8$.
\item Match of local density: we adopt a local density indicator $\rm N_{1Mpc}$, which is the count of galaxies brighter than $\rm M_{r} = -19.5$ and with redshifts differing less than $\rm 1000\;km\;s^{-1}$ from that of the target galaxy, in the surrounding sky area of radius = 1 Mpc (the count includes the target galaxy itself if it is brighter than $\rm M_{r} = -19.5$). By means of $\rm N_{1Mpc}$, we classify galaxies into 4 environmental categories: field ($\rm N_{1Mpc} \leq 3$), small group ($\rm 4 \leq N_{1Mpc} \leq 6$), large group ($\rm 7 \leq N_{1Mpc} \leq 10$), and cluster ($\rm N_{1Mpc}>10$). The control galaxy shall be in the same environmental category of the paired galaxy.
\item Has the closest redshift, among all qualified candidates, to that of paired galaxy.
\end{enumerate}

Finally, we have a control sample of 1320 (1167 unique) galaxies which are 10-to-1 matched to the 132 paired spiral galaxies in H-KPAIR. We allow galaxies to be included more than once in the control sample as long as there is no duplication among matches to any given paired galaxy. The $\rm M_{star}$ of control galaxies are estimated from the $\rm L_K$ and g-r color using the same method as that for paired galaxies (see Section \ref{sec:sample}).

\begin{figure}[htb!]
\includegraphics[width=0.47\textwidth]{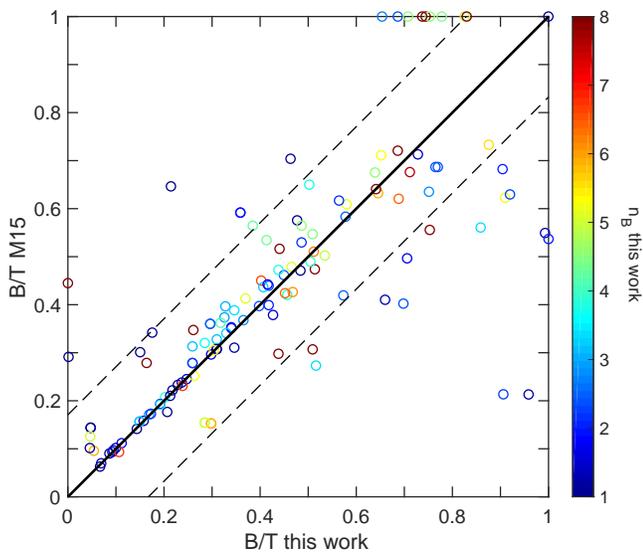}
\caption{The B/T ratios estimated in this work compared with those by \citet{2015MNRAS.446.3943M} for a subsample of control galaxies (see text). The solid line shows where the two results are identical with each other, and the dashed lines mark the 1$\sigma$ deviations. The color bar shows the S\'ersic index of the bulge component in the results of this work.
\label{fig:compare}}
\end{figure}

In order to check whether the B/T ratios obtained using our method and those estimated by \citet{2015MNRAS.446.3943M} are consistent with each other, we make the following comparison. For each paired galaxy we pick out from the 10 matching control galaxies the one with the smallest $\chi^2/\nu$ in the table of \citet{2015MNRAS.446.3943M} and derive its B/T ratio using the same method that we used for H-KPAIR galaxies. In Figure~\ref{fig:compare}, our results on the B/T ratios of the 132 such galaxies in the control sample are compared with the corresponding values in \citet{2015MNRAS.446.3943M}. A good overall agreement is found. The average difference between the two results is only 0.002 with a standard deviation of 0.169.

It is worth noting that, same as for paired galaxies, we assign $\rm B/T=0$ to all control galaxies with bulge S\'ersic index $n \leq 2$. However, there is a minor difference between paired and control samples in this treatment: For paired sample our \textsc{galfit} results give the $n$ values in real numbers, while for the control sample, whose \textsc{galfit} results are taken from \citet{2015MNRAS.446.3943M}, the $n$ values are reduced to integers. This means that some of the control galaxies with $n=2$ originally have the bulge S\'ersic index in the range of $2 \leq n < 2.5$. In order to constrain the uncertainties due to this minor mismatch between the paired and control samples, we test our results with an alternative control sample in which galaxies with $n < 2$ (instead of $n \leq 2$) are assigned $\rm B/T=0$. In this case, the treatment actually applies only to control galaxies with original bulge S\'ersic index $n < 1.5$. No significant difference is found in any of our results when this alternative control sample is used.

\begin{figure}[htb!]
\includegraphics[width=0.47\textwidth]{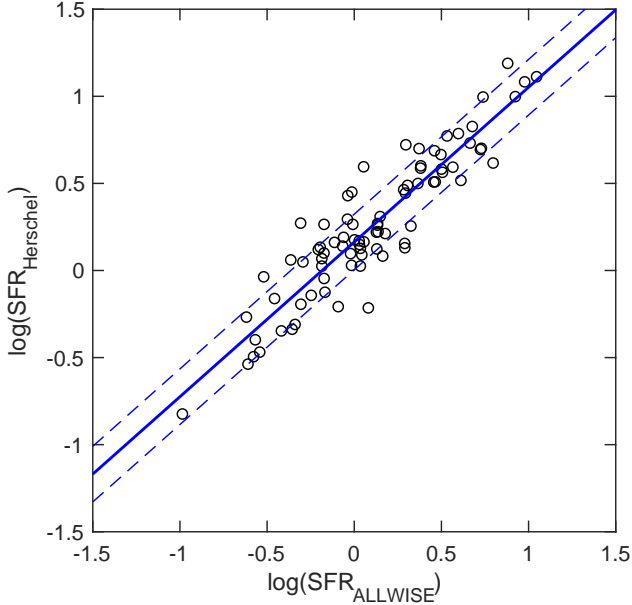}
\caption{Comparison between $\rm \log (SFR_{Herschel})$ and $\rm \log (SFR_{WISE})$ for control galaxies in \citet{2016ApJS..222...16C}. The solid line is the linear regression, and the dashed lines represent the 1$\sigma$ region.
\label{fig:her-wise}}
\end{figure}

The SFR of control galaxies is calculated from the Wide-field Infrared Survey Explorer \citep[WISE,][]{2010AJ....140.1868W,ALLWISE_data} w4-band $22\;\micron$ luminosity $\rm L_{22\micron}$, using the method given by \citet{2016ApJS..227....2S}. Namely we estimate the total IR luminosity by fit the luminosity-dependent IR templates of \citet*{2001ApJ...556..562C} to match the $\rm L_{22\micron}$, then use the conversion given by \citet{1998ApJ...498..541K}, adjusted to the Chabrier IMF using the 1.58 conversion factor \citep{2007ApJS..173..267S}. This yields the formula
\begin{equation}
\rm \log (SFR_{WISE}) = \log (L_{IR, CE}) - 9.966.
\end{equation}
The $\rm SFR_{WISE}$ so defined is different from the SFR of paired galaxies which are estimated using Herschel data \citep{2016ApJS..222...16C}. In order to make the two SFRs consistent with each other, we carried out the following analysis: Among the control galaxies of \citet{2016ApJS..222...16C}, which are one-to-one matched to H-KPAIR spirals, 82 galaxies have both Herschel and AllWISE $22\;\micron$ SFRs detection. We selected the H-KPAIR control sample \citep{2016ApJS..222...16C} instead of the pair members to avoid any pair blending issues in the $22\;\micron$ photometry. We carry out linear regression between $\rm SFR_{Herschel}$ and $\rm SFR_{WISE}$ of these galaxies, which is presented in Figure~\ref{fig:her-wise}. The result shows that
\begin{equation}
\rm \log(SFR_{Herschel}) = 0.89 \times \log(SFR_{WISE}) + 0.16.
\end{equation}
This conversion is applied to the $\rm SFR_{WISE}$ of our control galaxies to facilitate the comparison with paired galaxies. It is worth noting that the slightly non-linear relation between $\rm SFR_{Herschel}$ and $\rm SFR_{WISE}$ is likely due to the fact that the $22\;\micron$ emission in WISE w4-band is predominantly powered by massive ionizing stars \citep[$\rm > 10\; M_\sun$;][]{2007ApJ...666..870C} while the total IR luminosity used in the estimate of $\rm SFR_{Herschel}$ has significant contributions from intermediate and low mass stars and therefore decreases less rapidly when the massive star formation rate vanishes \citep{1996A&A...306...61B}. When a control galaxy has no detection in the WISE w4-band, the SFR upper limit is estimated from the upper limit of $\rm L_{22\micron}$.

\section{Dependence of \lowercase{s}SFR Enhancement on B/T Ratio} \label{sec:result}
\begin{figure}[htb!]
\includegraphics[width=0.47\textwidth]{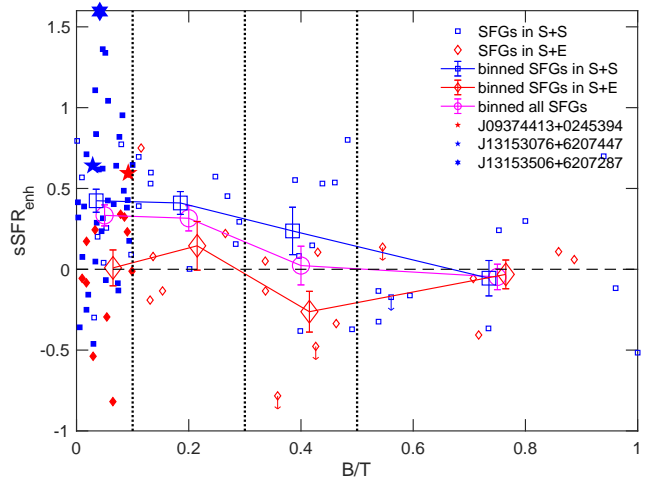}
\caption{Plot of $\rm sSFR_{enh}$ vs.~B/T for SFGs in S+S (small blue square) and in S+E (small red diamond) pairs. The means of $\rm sSFR_{enh}$ and their errors in the four B/T bins are plotted with large symbols and error bars, with black circles for SFGS in the total pair sample (including both S+S and S+E). The vertical dotted lines delineate the B/T bins, and the horizontal dashed line marks $\rm sSFR_{enh}=0$. Noting that all data points with $\rm B/T=0$ are plotted with a random offsets toward the positive direction of the \textit{x}-axis to avoid overlap, and are shown in filled markers. Three disk galaxies with NOEMA CO observations in \citet{2021ApJ...918...55X} are highlighted with pentagram and hexagram.
\label{fig:bt_ssfr}}
\end{figure}

For individual spiral galaxies in the pair sample, we define an sSFR enhancement index $\rm sSFR_{enh}$ as following:
\begin{equation}
\rm sSFR_{enh} = \log (sSFR_{pg}) - \log (sSFR_{med,ctrl}),
\end{equation}
where $\rm sSFR_{pg}$ is the sSFR of the paired spiral galaxy, and $\rm sSFR_{med,ctrl}$ the median of the sSFR of the 10 control galaxies. Following \citet{2016ApJS..222...16C}, spiral galaxies with $\rm \log (sSFR/yr^{-1}) < -11.3$ are regarded as in the red sequence and thus excluded from the analysis. The remaining 98 SFGs in the H-KPAIR sample are divided into 4 B/T bins: disky galaxies (B/T = 0–0.1), galaxies with small bulge (B/T = 0.1–0.3), galaxies with large bulge (B/T = 0.3–0.5), and bulge dominant galaxies (B/T = 0.5–1). Their sSFR enhancement is plotted against the B/T ratio in Figure~\ref{fig:bt_ssfr}. We use Kaplan-Meier (K-M) estimator \citep{KM1958,1985ApJ...293..192F} to take into account information in the upper limits. It shows that, for paired SFGs on the whole, there is a significant dependence of $\rm sSFR_{enh}$ on B/T ratio. In particular, only galaxies in the first two bins (with $\rm B/T<0.3$) have the average $\rm sSFR_{enh}$ significantly above zero while the average $\rm sSFR_{enh}$ of the last two bins (with $\rm B/T>0.3$) are consistent with no enhancement, supporting the hypothesis that large bulges suppress interaction-induced star formation \citep*{1996ApJ...464..641M}. Then, we divide the paired SFGs into S+S and S+E subsamples. For the SFGs in S+S pairs, the averages of $\rm sSFR_{enh}$ in the four B/T bins show a similar but stronger trend of anti-correlation as that for the total sample. Very strong sSFR enhancements ($\rm sSFR_{enh} > 0.7$~dex) are found almost exclusively in disky SFGs ($\rm B/T<0.1$) in S+S pairs. On the other hand, for the SFGs in S+E pairs, $\rm sSFR_{enh}$ vs B/T relation is rather flat, and none of the average $\rm sSFR_{enh}$ in individual bins is significantly above zero. Particularly, even in the first two B/T bins with small B/T ratios, where nearly half of SFGs in S+S pairs show strong enhancements ($\rm sSFR_{enh} \gtrsim 0.5$), SFGs in S+E pairs show no sSFR enhancement in general.

Star formation is fueled by cold gas, and sSFR can be decomposed into two terms $\rm sSFR=M_{gas}/M_{star}\times SFE_{gas}$, where $\rm SFE_{gas}=SFR/M_{gas}$ is the star formation efficiency of gas. \citet{2016ApJS..222...16C} derived the total gas mass ($\rm M_{gas}$) with a fixed gas-to-dust ratio of 100. They found significantly enhanced $\rm SFE_{gas}$ for SFGs in S+S pairs, but not for those in S+E pairs, compared to a control sample. On the other hand, they found no significant difference among the gas content ($\rm M_{gas}/M_{star}$) among SFGs in S+S, in S+E pairs, and in control sample.

\begin{figure}[htb!]
\includegraphics[width=0.47\textwidth]{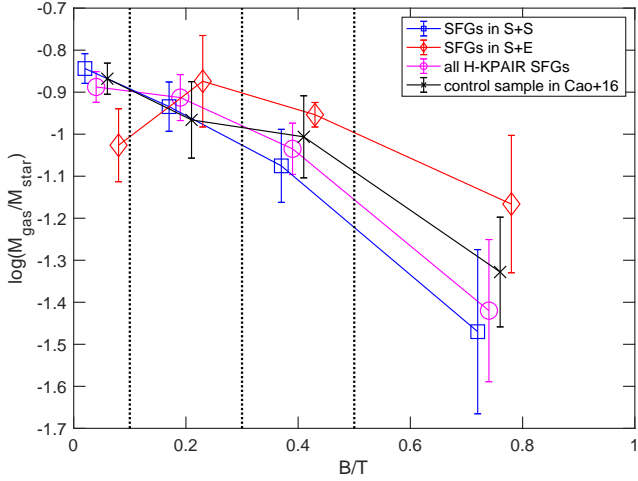}
\caption{Plot of means of $\rm \log (M_{gas}/M_{star})$ and errors in four B/T bins for SFGs in S+S (blue square), S+E (red diamond) pairs, total H-KPAIR (magenta circle), and control sample in \citet{2016ApJS..222...16C} (black cross).
\label{fig:gas-star}}
\end{figure}

In Figure~\ref{fig:gas-star} we show the means of the $\rm \log (M_{gas}/M_{star})$ in the four B/T ratio bins for 96 H-KPAIR SFGs with SFR detections, and of 95 SFR detected SFGs of the control sample in \citet{2016ApJS..222...16C}. Both the paired galaxies and the normal galaxies show a trend of the gas content decreasing with increasing B/T ratio. Consistent with \citet{2016ApJS..222...16C}, we find no significant difference among the gas contents of SFGs in S+S and S+E pairs and in control sample, in any of the B/T bins.

\begin{figure}[htb!]
\includegraphics[width=0.47\textwidth]{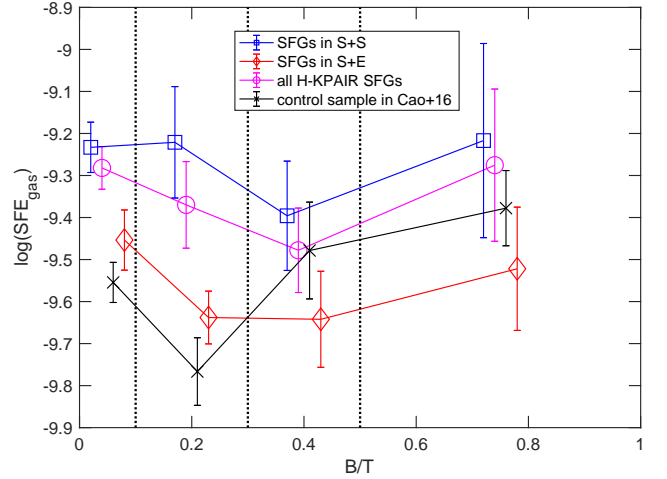}
\caption{Plot of means of $\rm \log (SFE_{gas})$ and errors in four B/T bins. The meaning of the symbols are the same as in Figure~\ref{fig:gas-star}.
\label{fig:sfe-gas}}
\end{figure}

In Figure~\ref{fig:sfe-gas} the means of $\rm \log (SFE_{gas})$ in four B/T bins are plotted for the same four samples. The SFGs in S+S pairs have systematically higher SFE than those in S+E pairs as \citet{2016ApJS..222...16C} showed, especially in the bin 1  ($\rm B/T<0.1$) and bin 2 ($\rm 0.1\leq B/T<0.3$), the differences are beyond 2$\sigma$. The enhanced SFE from SFGs in S+S pairs compared to that of control samples found in \citet{2016ApJS..222...16C} comes also mainly from these two bins ($\rm B/T<0.3$).

There are two kinds of cool gas in a galaxy: atomic gas and molecular gas. The connection of SFR with molecular gas is more direct than with atomic gas. \citet{2019A&A...627A.107L} carried out CO observation of 78 spiral galaxies in H-KPAIR and found that, compared with normal galaxies, paired SFGs in H-KPAIR sample show significant enhancement in $\rm M_{H_2}/M_{star}$ ratio but not in $\rm SFE_{H_2}$. When separated into S+S and S+E subsamples, SFGs in S+S pairs show higher $\rm M_{H_2}/M_{star}$ ($0.21\pm 0.11$~dex) and $\rm SFE_{H_2}$ ($0.18\pm 0.06$~dex) than those in S+E pairs.

\begin{figure}[htb!]
\includegraphics[width=0.47\textwidth]{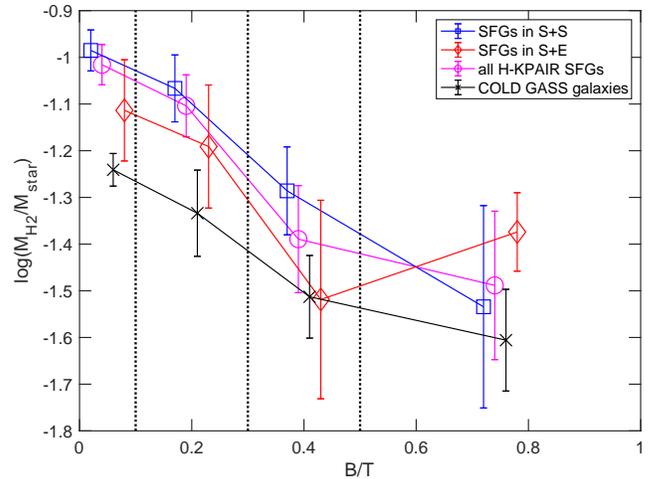}
\caption{Plot of means of $\rm \log (M_{H_2}-M_{star})$ and errors in four B/T bins for SFGs in S+S (blue square), S+E (red diamond) pairs, total H-KPAIR (magenta circle), and COLD GASS sample (black cross).
\label{fig:mol-star}}
\end{figure}

In Figure~\ref{fig:mol-star} we plot the means of the $\rm \log (M_{H_2}/M_{star})$ in the four B/T ratio bins for the 69 SFGs with WISE w4-band detection out of 78 H-KPAIR galaxies observed by \citet{2019A&A...627A.107L} in CO. Also plotted are the means of the $\rm M_{H_2}/M_{star}$ ratios of 93 normal SFGs with WISE w4-band detection that have CO data in the COLD GASS sample \citep{2011MNRAS.415...32S,2011MNRAS.415...61S} and B/T ratio data in \citet{2015MNRAS.446.3943M}. The plot shows a common trend for all samples that $\rm M_{H_2}/M_{star}$ decreases with increasing B/T. However, the trend for paired SFGs is steeper than that for normal galaxies, and paired SFGs with small B/T ratios have significant molecular fraction enhancement while those with $\rm B/T > 0.5$ have about the same molecular fraction as their counterparts in the normal galaxy sample. Also, spirals in S+E pairs show systematically lower $\rm M_{H_2}/M_{star}$ than those in S+S pairs, as was found previously by \citet{2019A&A...627A.107L}. The enhancement of $\rm M_{H_2}/M_{star}$ for disky ($\rm B/T<0.1$) SFGs in S+S pairs is 0.26~dex ($>4\sigma$).

\begin{figure}[htb!]
\includegraphics[width=0.47\textwidth]{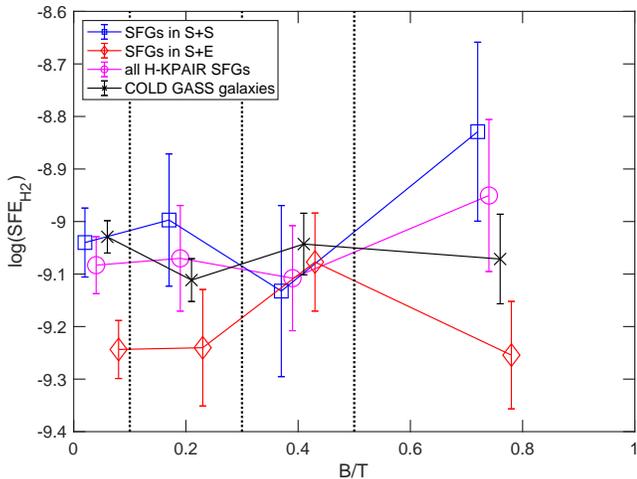}
\caption{Plot of means of $\rm SFE_{H_2}$ and errors in four B/T bins. The meaning of the symbols are the same as in Figure~\ref{fig:mol-star}.
\label{fig:sfe_mol}}
\end{figure}

In Figure~\ref{fig:sfe_mol} we show the means of $\rm \log (SFE_{H_2})$ in four B/T bins for the same four samples. It does not show significant trend for any sample and, compared to the single galaxies, no significant $\rm SFE_{H_2}$ enhancement for the paired SFGs is found in any B/T ratio bin. Most results for S+S and S+E subsamples have large errors because of relatively small sample sizes. The only noticeable difference is found in the first bin of $\rm B/T < 0.1$: the mean $\rm SFE_{H_2}$ of the disky SFGs in S+E pairs is $0.20\pm 0.09$~dex lower than that of their counterparts in S+S pairs.

\section{Discussion} \label{sec:discussion}
Our results show that indeed the sSFR enhancement is suppressed in paired SFGs with large bulges ($\rm B/T \geq 0.3$). In particular, very strong sSFR enhancement ($\rm sSFR_{enh} \gtrsim 0.7$) occurs almost exclusively in disky galaxies with $\rm B/T<0.1$. However, given the large scatter of the $\rm sSFR_{enh}$ versus B/T relation, the low frequency of interaction-induced starbursts cannot be explained solely by the B/T dependence. 

We find that spiral galaxies in S+E pairs have larger bulges than their counterparts in S+S pairs. This can explain partially the $\rm sSFR_{enh}$ difference between SFGs in S+E and S+S pairs found by \citet{2010ApJ...713..330X} and \citet{2016ApJS..222...16C}. But it cannot explain why in a same B/T bin the $\rm sSFR_{enh}$ of SFGs in S+E pairs are systematically lower than those in S+S pairs (Figure~\ref{fig:bt_ssfr}). Particularly, in the bin of disky galaxies ($\rm B/T<0.1$), SFGs in S+E show significantly lower $\rm M_{H_2}/M_{star}$ and $\rm SFE_{H_2}$ than their counterparts in the S+S subsample, and none of them has 
$\rm sSFR_{enh} > 0.7$ while only one has $\rm sSFR_{enh} > 0.5$. 

Why do some paired disky SFGs have strong sSFR enhancements while others do not, especially those in S+E pairs? According to \citet{2021ApJ...918...55X}, the systematically higher $\rm SFE_{H_2}$ of SFGs in S+S pairs than those in S+E pairs may be explained by the following scenario: The former may have higher chance to be in low-speed co-planer interactions which can trigger strong nuclear starbursts by tidal torques \citep{*1996ApJ...471..115B,2009ApJ...691.1168H}, while the latter are more likely in higher speed, higher incline angle interactions which tend to trigger ring galaxies that have more extended star formation and lower $\rm SFE_{H_2}$, such as what is observed in the S+E pair Arp~142 (its spiral component has $\rm B/T<0.1$). This hypotheses is based on their results that on average S+S pairs have lower local density and lower relative velocity than S+E pairs, therefore are more likely found in the field environment which favors co-planer interactions \citep{2014MNRAS.444.1453D}. On the other hand, S+E pairs are more likely found in groups/clusters where the interaction orbits are severely disturbed and randomly oriented. More investigations are needed to explore whether this mechanism can explain the difference in the $\rm sSFR_{enh}$ of disky SFGs in S+S and S+E pairs.

\section{Conclusion} \label{sec:conclusion}
In this paper we present a bulge-disk decomposition catalog of a 2MASS Ks-band-selected close major-merger galaxy pair sample (H-KPAIR). The decompositions are derived by two-dimensional, two-component fits on SDSS r-band images using \textsc{galfit}. With this catalog and a control sample of single galaxies selected from a large catalog of SDSS galaxies with \textsc{galfit} results \citep{2015MNRAS.446.3943M}, we are able to study the dependence of the interaction-induced sSFR enhancement on the B/T ratio, and verify the theoretical prediction that large bulges can suppress the star formation enhancement in interacting galaxies \citep{*1996ApJ...464..641M}. We also investigate the effects of the B/T ratio dependence of the sSFR enhancement on the differences between star-forming galaxies (SFGs) in spiral+spiral (S+S) pairs and spiral+elliptical (S+E) pairs. Our main results are:

\begin{enumerate}
\item There is a strong monotonic dependence of sSFR enhancement on the B/T ratio, in the sense that sSFR enhancement decreases with increasing B/T. On average, only paired SFGs with $\rm B/T<0.3$ show significant sSFR enhancement.
\item When separated into S+S and S+E subsamples, 
the S+S subsample shows a similar (albeit slightly stronger) trend. Very strong sSFR enhancements ($\rm sSFR_{enh} > 0.7$~dex) are found almost exclusively in disky SFGs ($\rm B/T<0.1$) in S+S pairs. 
\item However, for SFGs in S+E pairs, the $\rm sSFR_{enh}$ does not show any clear B/T dependence, nor any significant enhancement in any B/T bin. The mean $\rm sSFR_{enh}$ of SFGs in S+E pairs is lower than that of their counterparts in S+S pairs in all B/T bins.
\item Spiral galaxies in S+E pairs have higher B/T ratios than their counterparts in S+S pairs, with means of $\rm B/T = 0.35\pm 0.05$ and $\rm B/T=0.26\pm 0.03$, respectively.
\item Total gas content($\rm M_{gas}/M_{star}$) and molecular gas content ($\rm M_{H_2}/M_{star}$) are both anti-correlated with B/T for SFGs in both kinds (S+S and S+E) of pairs, and for single SFGs. Disky ($\rm B/T<0.1$) SFGs in S+S pairs show Significant ($>4\sigma$) $\rm M_{H_2}/M_{star}$ enhancement. SFGs in S+E pairs have systematically lower $\rm M_{H_2}/M_{star}$ than their counterparts in S+S pairs, and show no significant $\rm M_{H_2}/M_{star}$ enhancement in any B/T bin. No significant difference of $\rm M_{gas}/M_{star}$ is found between both kinds of pairs, and no $\rm M_{gas}/M_{star}$ enhance of paired galaxies is found in any B/T bin.
\item It appears that both the star formation efficiency calculated by total gas ($\rm SFE_{gas} = SFR/M_{gas}$) and by molecular gas ($\rm SFE_{H_2} = SFR/M_{H_2}$) do not depend on the B/T ratio. There is significant $\rm SFE_{gas}$ enhancement for paired SFGs with $\rm B/T<0.3$, and the enhancement is mainly from the SFGs in S+S pairs. No significant $\rm SFE_{H_2}$ enhancement is found for paired SFGs in any B/T bin. The $\rm SFE_{H_2}$ of the disky SFGs ($\rm B/T<0.1$) in S+E pairs is on average $0.20\pm 0.09$~dex lower than that of their counterparts in S+S pairs.
\end{enumerate}

\begin{acknowledgments}

This work is supported by the National Key R\&D Program of China No. 2017YFA0402704 and by National Natural Science Foundation of China (NSFC) No. 11873055 and No.11933003, and is sponsored (in part) by the Chinese Academy of Sciences (CAS) through a grant to the CAS South America Center for Astronomy (CASSACA). CKX acknowledges NSFC grants No. 11733006.

The authors would like to thank Gaoxiang Jin, Y.Sophia Dai, Cheng Cheng, Hai Xu, Piaoran Liang, Zijian Li, Dandan Xu, Shengdong Lu, Hongming Tang, Yiru Chen and Yilun Wang for their helpful discussions.

Funding for the Sloan Digital Sky Survey IV has been provided by the Alfred P. Sloan Foundation, the U.S. Department of Energy Office of Science, and the Participating Institutions. SDSS acknowledges support and resources from the Center for High-Performance Computing at the University of Utah. The SDSS web site is www.sdss.org.

SDSS is managed by the Astrophysical Research Consortium for the Participating Institutions of the SDSS Collaboration including the Brazilian Participation Group, the Carnegie Institution for Science, Carnegie Mellon University, Center for Astrophysics | Harvard \& Smithsonian (CfA), the Chilean Participation Group, the French Participation Group, Instituto de Astrofísica de Canarias, The Johns Hopkins University, Kavli Institute for the Physics and Mathematics of the Universe (IPMU) / University of Tokyo, the Korean Participation Group, Lawrence Berkeley National Laboratory, Leibniz Institut für Astrophysik Potsdam (AIP), Max-Planck-Institut für Astronomie (MPIA Heidelberg), Max-Planck-Institut für Astrophysik (MPA Garching), Max-Planck-Institut für Extraterrestrische Physik (MPE), National Astronomical Observatories of China, New Mexico State University, New York University, University of Notre Dame, Observatório Nacional / MCTI, The Ohio State University, Pennsylvania State University, Shanghai Astronomical Observatory, United Kingdom Participation Group, Universidad Nacional Autónoma de México, University of Arizona, University of Colorado Boulder, University of Oxford, University of Portsmouth, University of Utah, University of Virginia, University of Washington, University of Wisconsin, Vanderbilt University, and Yale University.

This publication makes use of data products from the Two Micron All Sky Survey, which is a joint project of the University of Massachusetts and the Infrared Processing and Analysis Center/California Institute of Technology, funded by the National Aeronautics and Space Administration and the National Science Foundation.

This publication makes use of data products from the Wide-field Infrared Survey Explorer, which is a joint project of the University of California, Los Angeles, and the Jet Propulsion Laboratory/California Institute of Technology, funded by the National Aeronautics and Space Administration.

\end{acknowledgments}

\bibliography{sample631}{}
\bibliographystyle{aasjournal}

% Include this line if you are using the \added, \replaced, \deleted
% commands to see a summary list of all changes at the end of the article.
\listofchanges

\end{document}